\documentclass[preprint,sort&compress,12pt]{elsarticle}
\usepackage{bbm}
\usepackage{amstext}
\usepackage{amsmath}
\usepackage{amssymb}
\usepackage{mathrsfs}
\usepackage{stmaryrd}
\usepackage{bm}
\usepackage{pstricks}
\usepackage{pst-coil}
\usepackage{pst-3d}
\usepackage{amsfonts}
\usepackage{amsthm}
\usepackage{CJK}

\def\journal{
 }
\def\ftimes{\hfill Completed on January 15, 2019}

\newcommand{\mm}{\mathrm}
\newcommand{\ml}{\mathcal}

\newcommand{\be}{\begin{equation}}
\newcommand{\bea}{\begin{equation}\begin{aligned}}
\newcommand{\beas}{\begin{equation*}\begin{aligned}}
\newcommand{\eeas}{\end{aligned}\end{equation*}}
\newcommand{\eea}{\end{aligned}\end{equation}}
\newcommand{\ee}{\end{equation}}

\renewcommand{\div}{{\rm div }}

\begin{document}
\begin{CJK*}{GBK}{song}
\begin{frontmatter}
\title{
On  Classical Solutions of Rayleigh--Taylor Instability in Inhomogeneous\\
Incompressible Viscous Fluids in Bounded Domains}

\author[FJ]{Fei Jiang}
\ead{jiangfei0591@163.com}
\author[FJ]{Youyi Zhao}
\ead{zhaoyouyi957@163.com}
\address[FJ]{College of Mathematics and
Computer Science, Fuzhou University, Fuzhou,   China.}

\begin{abstract}
We study the existence of unstable classical solutions of the Rayleigh--Taylor instability problem (abbr. RT problem) of an inhomogeneous incompressible viscous fluid in a bounded domain. We find that, by using an existence theory of (steady) Stokes problem and an iterative technique, the initial data of classical solutions of the linearized RT problem can be modified to new initial data, which can generate local-in-time classical solutions of the RT problem, and are close to the original initial data. Thus, we can use a classical bootstrap instability method to further obtain  classical solutions of (nonlinear) RT instability based on the ones of linear RT instability.
\end{abstract}
\end{frontmatter}


\newtheorem{thm}{Theorem}[section]
\newtheorem{lem}{Lemma}[section]
\newtheorem{pro}{Proposition}[section]
\newtheorem{concl}{Conclusion}[section]
\newtheorem{cor}{Corollary}[section]
\newproof{pf}{Proof}
\newdefinition{rem}{Remark}[section]
\newtheorem{definition}{Definition}[section]

\section{Introduction}\label{introud}
\numberwithin{equation}{section}

The motion of a three-dimensional (3D) nonhomogeneous incompressible
viscous fluid in the presence of a uniform gravitational field in a
bounded domain $\Omega\subset {\mathbb R}^3$ is governed by the following
Navier--Stokes equations:
\begin{equation}\label{0101xx}\left\{\begin{array}{l}
 \rho_t+{  v}\cdot\nabla \rho=0,\\[1mm]
\rho v _t+\rho {  v}\cdot\nabla {  v}+\nabla p=\mu\Delta v -g\rho {e}_3,\\[1mm]
\mathrm{div} v =0,\end{array}\right.\end{equation}
where the unknowns $\rho:=\rho(x,t)$, $ v := v (x,t)$ and $p:=p(x,t)$ denote the density, velocity and pressure of the fluid, respectively; $\mu>0$ stands for the coefficient of shear viscosity, $g>0$ for the gravitational constant, $ {e}_3=(0,0,1)$ for the vertical unit vector, and $-g {e}_3$ for the gravitational force.
In the system \eqref{0101xx} the equation \eqref{0101xx}$_1$ describes the mass conservation, \eqref{0101xx}$_2$   the balance law of momentum, and \eqref{0101xx}$_3$ the incompressible condition.

The stability/instability of viscous incompressible flows governed by the Navier--Stokes equations is a classical subject with a very extensive literature over more than 100 years \cite{CSHHSCPO,FSPNSPR}. In this paper we study  the Rayleigh--Taylor (RT) instability  to the system \eqref{0101xx}.  To this purpose, we consider a density profile
$\bar{\rho}:=\bar{\rho}(x_3)\in  C^1(\bar{\Omega})$, which satisfies
\begin{eqnarray}\label{0102}
\inf_{ {x}\in
 \Omega}\{\bar{\rho}( {x})\}>0,\end{eqnarray}
 and an RT condition
\begin{eqnarray}\label{0103}\bar{\rho}'( {x}_3^0)>0 \mbox{ for some } x^0\in  \Omega,
\end{eqnarray}
where $x^0_3$ denotes the third component of $x^0$. Then we further define
a pressure profile $\bar{p}$ (unique up to a constant) by the relation
$$\nabla \bar{p}=-g\bar{\rho} e_3.$$
The condition \eqref{0103} means that there is a region in which the RT density profile has
larger density with increasing $x_3$ (height), thus leading to the RT instability.

Obviously, $(\rho,v,p)=(\bar{\rho},0,\bar{p})$ is an RT equilibrium-state solution of the system \eqref{0101xx}. Now, we denote the perturbation by
$$ \varrho=\rho -\bar{\rho},\quad  {u}= v - {0},\quad q=p-\bar{p},$$
then, $(\varrho , {u},q)$ satisfies the perturbation equations:
\begin{equation}\label{0105}\left\{\begin{array}{l}
\varrho_t+{  u}\cdot\nabla (\varrho+\bar{\rho})=0, \\[1mm]
(\varrho+\bar{\rho}){  u}_t+(\varrho+\bar{\rho}){  u}\cdot\nabla
{  u}+\nabla q=\mu\Delta{ {{u}}}-g\varrho {e}_3,\\[1mm]
 \mathrm{div} {u}=0.\end{array}\right.  \end{equation}
We specify the
initial-boundary value conditions:
\begin{equation}
(\varrho,{  u})|_{t=0}=(\varrho_0,{  u}_0),\ \label{0107}
{  u}|_{\partial\Omega}={  0}.
\end{equation}
The above initial-boundary value problem \eqref{0105}--\eqref{0107} is called the (nonlinear) RT (instability) problem.  The initial-boundary value  problem obtained by omitting the nonlinear terms in RT problem is called the linearized RT problem.

The RT instability is well known as gravity-driven instability in fluids when a heavy fluid is on top of a light one. Instability of the linearized problem of \eqref{0105}--\eqref{0107} (i.e. linear instability) was first introduced by Rayleigh in 1883 \cite{RLAP}. Similar results of linear RT instability were established for two layer incompressible/compressible fluids with a free interface (so-called stratified fluids), where the RT equilibrium-state solution is a denser fluid lying above a lighter one separated by an interface, please refer to \cite{GYTI1,GYTI2,CSHHSCPO,wang2011viscous,JJTIWYHCMPxxx} for relevant progress.

In 2003, Hwang and Guo \cite{HHJGY} proved the existence of classical solutions of (nonlinear)  RT instability in the sense of $L^2$-norm for a 2D nonhomogeneous incompressible inviscid fluid (i.e. $\mu=0$ in \eqref{0105}), where the classical solution defined on $\mathbb{R}^2$ is periodic in the horizontal direction. Then Jiang--Jiang further showed the existence of strong solutions of RT instability for the RT problem \eqref{0105}--\eqref{0107} in the sense of $L^2$-norm \cite{JFJSO2014}. Similar results of RT instability were established for stratified incompressible viscous fluids, see \cite{PJSGOI5x,WMRTITNS2017,JFJSOMITNWZ} for instance.

To our best knowledge,  all obtained  solutions of (nonlinear)  RT  instability with specific boundary-value conditions  in \cite{JFJSO2014,PJSGOI5x,WMRTITNS2017,JFJSOMITNWZ} are strong solutions. Moreover there are also not available results for the existence of classical solutions of other flow instabilities in incompressible fluids with specific boundary-value conditions. Compared with strong solutions, the initial data of classical solutions satisfy more compatibility conditions. This results in that the both linearized and nonlinear problems often have different compatibility conditions, and thus
the initial data of classical solutions of linearized problem  can not directly generate local-in-time unique classical solutions for the nonlinear problem in the classical bootstrap instability method, i.e., a method of constructing nonlinear instability solutions based on linear instability solutions \cite{GYHCSDDC,GYYQH,GYSWIC,GYSWICNonlinea}.
 This is the main reason that Jiang--Jiang \cite{JFJSO2014} considered the existence of strong solutions of RT instability for the RT problem \eqref{0105}--\eqref{0107}, since
 the both linearized and nonlinear RT problems have the same compatibility conditions under the case of the strong solutions.  Of course, the incompatibility problem also does not appears, if the fluid domain is periodic or  whole space. Hence a lot of mathematical results for unstable classical solutions of various flow instability problems defined on periodic domain or whole space can easily be founded.

 In this paper, we further find that, by using existence theory of Stokes problem and an iterative technique, the initial data of unstable classical solutions of the linearized RT problem can be modified new initial data, so that the obtained new initial data can generate local-in-time unique classical solutions for (nonlinear) RT problem, and are close to the original initial data. Thus, we can use a classical bootstrap instability method, developed by Guo et.al. \cite{GYHCSDDC}, to obtain  unstable classical solutions of the RT problem. Our method can be applied to other relevant instability problems. In particular, we will take a magnetic RT problem as an example in Section \ref{201812311629}.

\subsection{Main results}

Before stating our main result, we shall introduce some mathematical notations of Sobolev spaces.
\begin{align}
&L^p (D)=W^{0,p}(D),\  {H}^i(D):=W^{i,2}(D) ,\nonumber \\
&H^1_0(D):=\{w\in {H}^1(D )~|~w|_{\partial D}=0\mbox{ in the sense of trace}\},\nonumber \\
  &H_0^i(D):=H_0^1(D) \cap H^i(D),\  H^1_\sigma(D):=\{w\in H^1_0(D)~|~\div w=0\mbox{ in }D\}, \nonumber\\
 &H^i_\sigma(D):=H^1_\sigma(D) \cap H^i(D),\
  \underline{H}^i(D):=\left\{w\in {H}^i(D)~\left|~\int_D w\mm{d}y=0\right\}\right., \nonumber
\end{align}
where $1\leqslant  p\leqslant \infty$   and   $i\geqslant 0$ are integers. In addition, $I_T:=(0,T)$ and $u_{\mm{h}}:=(u_1,u_2)$.
Next we state our main result.

\begin{thm}\label{thm:0102xx}
Let $\Omega$ be a $C^5$-bounded domain, and
the density profile $\bar{\rho}\in C^5(\overline\Omega)$ satisfies \eqref{0102} and \eqref{0103}.
 Then, a zero solution to the RT problem \eqref{0105}--\eqref{0107}
 is unstable in the Hadamard sense, that is, there are positive constants $\Lambda$, $m_0$, $\varepsilon$ and $\delta_0$,
 and
$$(\tilde{\varrho}^0,\tilde{ {u}}^0,\tilde{q}^0,u^{\mm{r}},q^{\mm{r}})\in H^4_0(\Omega)\times H^4_\sigma(\Omega)\times \underline{H}^3(\Omega)\times H^4_\sigma(\Omega)\times \underline{H}^3(\Omega), $$
such that, for any given  $\delta\in (0,\delta_0)$,
there is a unique classical solution $({\varrho}, {u})\in C^0(\overline{I_T },H^4(\Omega)\times H_\sigma^4(\Omega))$, with a unique associated (perturbation) pressure $q\in C^0(\overline{I_T},\underline{H}^3)$, to the RT problem with the initial data
 $$(\varrho^0, {u}^0,q^0):= \delta (\tilde{\varrho}^0,\tilde{u}^0,\tilde{q}^0)+\delta^0(0, \tilde{u}^{\mm{r}},\tilde{q}^{\mm{r}}),$$
but the solution satisfies
\begin{align}
& \|\varrho(T^\delta)\|_{L^1(\Omega)},\ \|u_{\mm{h}}(T^\delta)\|_{L^1(\Omega)},\ \|{u}_3(T^\delta)\|_{L^1(\Omega)} \geqslant {\varepsilon}\label{0115}
\end{align}
for some escape time $T^\delta:=\frac{1}{\Lambda}\mm{ln}\frac{2\varepsilon}{m_0\delta}\in
 {I_T }$. In addition, the initial data $(\varrho^0,u^0,q^0)$ satisfies compatibility conditions:
\begin{alignat}{2}
&\mm{div}\left(\left( u^0\cdot\nabla
 u^0+( \nabla q^0-\mu\Delta u^0+g\varrho^0 {e}_3\right)/(\varrho^0+\bar{\rho})\right)=0& &\ \mbox{ in }\Omega,\label{201809101928XXx}\\
& \nabla q^0-\mu\Delta u^0+g\varrho^0 {e}_3=0 & &\ \mbox{ on }\partial\Omega. \label{xx201809101926xxsdfsXXX}
\end{alignat}
\end{thm}

The proof of Theorem \ref{thm:0102xx} is based on a  bootstrap instability method, which has its origin in \cite{GYSWIC,GYSWICNonlinea}. Later, various versions of bootstrap approaches were established by many authors, see \cite{FSSWVMNA,GYHCSDDC,FSNPVVNC,FrrVishikM} for examples. In this paper, we adapt the version of bootstrap instability method in \cite[Lemma 1.1]{GYHCSDDC} to prove Theorem \ref{thm:0102xx} by further introducing some mathematical techniques. Next we briefly sketch the proof procedure.

Firstly, we use energy method to derive that the local-in-time solutions of the (nonlinear) RT problem enjoy  a  Gronwall-type energy inequality, see Proposition \ref{pro:0401xx}. Secondly, we introduce unstable solutions to the linearized RT problem, see Proposition \ref{thm:0101xx},  and then use an existence theory of Stokes problem and an iterative technique to modify initial data of solutions of the linearized RT problem, so that the obtained modified initial data  satisfy the compatibility conditions  \eqref{201809101928XXx}--\eqref{xx201809101926xxsdfsXXX}, and are close to the original initial data, see Propositions \ref{lem:modfiedxx}. Finally, we introduce the error estimates between the solutions of the linearized and nonlinear  RT problems, and then prove the existence of escape time $T^\delta$. The detailed proof of Theorem \ref{thm:0102xx} will be provided in Sections  \ref{201806101821}.

\subsection{Extensions}\label{201812311629}

Recently, the existence results of strong solutions of RT instability are obtained in the both inhomogeneous  incompressible  viscoelastic fluid and inhomogeneous incompressible  viscous magnetohydrodynamic (MHD) fluid without resistivity \cite{JFJWGCOSdd,JFJSJMFMOSERT}.
By modifying the proof of Theorem \ref{thm:0102xx}, we can further obtain classical solutions of RT instability in the sense of $L^1$-norm in the both fluids. Next we take the inhomogeneous incompressible  viscous MHD fluid without resistivity  as an example.

To begin with, let us recall  the equations of an inhomogeneous incompressible  viscous MHD fluid without resistivity  in the presence of a uniform gravitational field:
\begin{equation}\label{0101}\left\{\begin{array}{l}
 \rho_t+ v\cdot \nabla \rho =0,\\[1mm]
\rho v_t+\rho v\cdot\nabla v + \nabla (p+\lambda  |M |^2/2)-\mu \Delta v= \lambda M \cdot\nabla M-\rho g e_3, \\[1mm]
M_t=M\cdot \nabla v-v\cdot\nabla M, \\[1mm]
\mathrm{div} M=0,\\
\mathrm{div} {v}=0,\end{array}\right.\end{equation}
where $M$ represents the magnetic field, and  $\lambda$  the permeability of vacuum divided by $4\pi$. The equation \eqref{0101}$_2$ describes the motion of the MHD fluids driven by the gravitational field along the negative $x_3$-direction.  The equation  \eqref{0101}$_3$ is called the induction equation, and the equation \eqref{0101}$_4$ represents that magnetic field is source-free.

As in \cite{JFJSJMFMOSERT},
we consider that the motion of the MHD fluid is horizontally periodic in a layer domain. So we shall introduce a horizontally periodic domain with finite height, i.e.
$$\Omega_{\mm{p}}:=\mathbb{T}\times (0,h),$$
where $\mathbb{T}:=\mathbb{T}_1\times \mathbb{T}_2$, $ \mathbb{T}_i=2\pi L_i(\mathbb{R}/\mathbb{Z})$, and $2\pi L_i$ ($i=1$, $2$) are the periodicity lengths.
We denote $\mathbb{T}\times \{0,h\} $ by $\partial\Omega_{\mm{p}}$.

For a given constant vector $\bar{M}$ and a density profile $\bar{\rho}\in C^1(\overline{\Omega_{\mm{p}}})$ satisfying \eqref{0102}--\eqref{0103} with $\Omega_{\mm{p}}$ in place of $\Omega$, the solution $(\rho ,v, M)=(\bar{\rho},0,\bar{M})$ defines an MRT equilibrium state to the system \eqref{0101} with an associated pressure profile $\bar{p}$ (unique up to a constant) satisfying
$\nabla \bar{p}=-\bar{\rho}g  {e}_3$.

Denoting the perturbation around the MRT equilibrium state by
$$ \varrho=\rho -\bar{\rho},\quad  {v}= {v}- {0},\quad N= M -\bar{M},\quad  q=p+ \lambda |M|^2/2-\bar{p},$$
then $(\varrho , {v},N,q )$ satisfies the perturbation  equations
\begin{equation}\label{0103xx}
\left\{\begin{array}{l}  \varrho_t+{  v}\cdot\nabla (\varrho+\bar{\rho})=0, \\[1mm]
(\varrho+\bar{\rho}){  v}_t+(\varrho+\bar{\rho}){  v}\cdot\nabla
{ v}+\nabla q -\mu \Delta v  =\lambda (N+\bar{M})\cdot \nabla N - \varrho  g e_3,
\\[1mm]
N_t=(N+\bar{M})\cdot \nabla v- v\cdot\nabla N,\\[1mm]
\mathrm{div}v= \mathrm{div} {N}=0.\end{array}\right.  \end{equation}
We specify the initial-boundary value  conditions:
\begin{equation}
(\varrho ,v,N)|_{t=0}=(\varrho^0,v^0,N^0)\mbox{ and } v|_{\partial\Omega_{\mm{p}}} ={  0} .
\label{0105xxx}
\end{equation}
The initial boundary value problem \eqref{0103xx}--\eqref{0105xxx} is called the  MRT problem.

There is a vast literature on linear RT instability in MHD fluids (without resistivity), see \cite{KMSMSP,HRWP,CSHHSCPO,DRJFJS,WYC,JFJSWWWOA,WYTIVNMI} for instance. Here we only mention the mathematical progress in the nonlinear RT instability.
Hwang probably first prove the existence of classical solutions of RT instability in the inhomogeneous incompressible inviscid  MHD fluid defined on a periodic domain \cite{HHVQ}. Then, Jiang--Jiang further proved the the existence of classical solutions of RT instability  for the system \eqref{0103xx} defined on the domain $\mathbb{T}\times \mathbb{R}$ \cite{JFJSWWWN}. Recently, Jiang--Jiang \cite{JFJWGCOSdd} further obtained strong solutions of RT instability for the MRT problem \eqref{0103xx}--\eqref{0105xxx} with $\bar{M}=(0,0,\bar{M}_3)$ under the instability condition
\begin{equation}
\label{201812291920}
|\bar{M}_3|<m_{\mm{c}} :=\sup_{0\neq w\in H_{\sigma}^1(\Omega_{\mm{p}})}{\frac{g\int_{\Omega_{\mm{p}}}\bar{\rho}' w_3^2\mm{d}y}{\lambda\|\partial_3 w\|^2_{L^2(\Omega_{\mm{p}})}}}.
\end{equation}
Similar instability results can be also found in the incompressible/compressible stratified MHD fluids \cite{JFJSOMITNWZ,JFJSNS}. It should be noted that the solutions of RT instability constructed in \cite{JFJWGCOSdd,JFJSOMITNWZ,JFJSNS} are strong solutions. In this paper, we further obtain  classical solutions of RT instability to the MRT problem.

\begin{thm}\label{201806dsaff012301}
Let the density profile $\bar{\rho}\in C^6(\overline{\Omega})$ satisfy \eqref{0102} and \eqref{0103}, and  $\bar{M}$ be a constant vector. We further assume that  $\bar{\rho}$ and $\bar{M}$ satisfy  \eqref{201812291920}. Then  a zero solution of the MRT problem   \eqref{0103xx}--\eqref{0105xxx} is unstable in the Hadamard sense, that is, there are positive constants $C$ and $\delta_0$ such that, for any $\delta\in (0,\delta_0)$, there exist $(\varrho^0,v^0,N^0,q ^0 )\in H^4(\Omega_{\mm{p}}) \times  H^4_\sigma(\Omega_{\mm{p}}) \times \underline{H}^3(\Omega_{\mm{p}})$ satisfying
\begin{equation}
\| ({\varrho}^0, v^0,N^0)\|_{H^4(\Omega_{\mm{p}})}+\|q ^0\|_{H^3(\Omega_{\mm{p}})}\leqslant C\delta,
\end{equation}
and  a unique classical solution $(\varrho,v,N,q )\in C^0(\overline{I_T},H^4(\Omega_{\mm{p}})\times  H^4_\sigma(\Omega_{\mm{p}}) \times H^4(\Omega_{\mm{p}}) \times \underline{H}^3(\Omega_{\mm{p}}))$ to the MRT problem with  initial data $ ({\varrho}^0, v^0,N^0,q ^0)  $,
but the solution satisfies
\begin{align}
&\|v_{\mm{h}}(T^\delta)\|_{L^1(\Omega_{\mm{p}})},\  \|v_3(T^\delta)\|_{L^1(\Omega_{\mm{p}})}\geqslant \varepsilon ,\label{201809041319}\\
& \|N_{\mm{h}}(T^\delta)\|_{L^1(\Omega_{\mm{p}})},\ \|N_3(T^\delta)\|_{L^1(\Omega_{\mm{p}})}\geqslant  \varepsilon, \mbox{ if }
\bar{M}_3\neq0  \label{201809041319xx}
\end{align}
for some escape time $T^\delta \in I_T$. Moreover, the initial data $(\varrho^0,v^0,N^0,q ^0)$ satisfies necessary compatibility  conditions
\begin{alignat}{2}
& \mm{div} N^0=0, \label{20180923221611xx}\\
& \mm{div} ( v^0 \cdot \nabla v^0 +   (\nabla q ^0-\mu\Delta  v^0 & &
\\
&-\lambda (N^0+\bar{M})\cdot \nabla N^0 +g\varrho^0 e_3) / (\varrho^0+\bar{\rho})  )=0 & &\ \mbox{ in }\Omega_{\mm{p}},  \label{2018092322161}  \\
&  \nabla q ^0-\mu\Delta  v^0 -\lambda (N^0+\bar{M})\cdot \nabla N^0 +  g  \varrho^0 e_3   =0 & &\ \mbox{ on }\partial\Omega_{\mm{p}}.\label{20180923221611}
\end{alignat}
\end{thm}

Here we briefly introduce the proof procedure of Theorem \ref{201806dsaff012301}.
First we shall rewrite the MRT problem \eqref{0103xx}--\eqref{0105xxx} in Lagrangian coordinates,  thus getting a so-called transformed MRT problem, which enjoys fine mathematical structure  so that we can establish the desired Gronwall-type energy inequality in Lagrangian coordinates. Then we can prove the existence of classical solutions of RT instability for the transformed MRT problem by following the proof frame of Theorem \ref{thm:0102xx} with finer mathematical analysis. Finally, by an inverse transformation of Lagrangian coordinates, we obtain Theorem \ref{201806dsaff012301}.

The rest two sections are devoted to the proofs of Theorems \ref{thm:0102xx}--\ref{201806dsaff012301}, resp..

\section{Proof of Theorem \ref{thm:0102xx}}\label{201806101821}
This section is devoted to the proof of Theorem \ref{thm:0102xx}.
We shall introduce some simplified notations throughout this section.
$$
\begin{aligned}
& \int:=\int_\Omega,\ \|\cdot \|_{i} :=\|\cdot \|_{H^{i}(\Omega)},\ \|(u,q)\|_{\mm{S},k}:=\sqrt{\|u\|_{k+2}^2+\|   q\|_{k+1}^2 } ,\\
& \mathcal{E}(t):=\sum_{i=0}^2\|\partial_t^i\varrho\|_{4-i}^2+
\sum_{i=0}^1\|\partial_t^i(u,q)\|_{\mm{S},2-2i}^2 +\|u_{tt}\|_0^2,\\
  & \mathcal{D}(t):=\sum_{i=0}^1\|\partial_t^i(u,q)\|_{\mm{S},3-2i}^2 +\|u_{tt}\|_1^2. \end{aligned} $$
In addition, the function space $X(\Omega)$ is simplified by $X$ for $X=L^p$, $H^i$ and so on, and $a\lesssim b$ means that $a\leqslant cb$ for some constant $c>0$, which may depend on the domain $\Omega$,  the density profile $\bar{\rho}$ and physical parameters $\mu $ and  $g$, and may vary from line to line.

\subsection{Gronwall-type energy inequality of nonlinear solutions}\label{201805081731}

We derive that any small solution of the RT problem enjoys a Gronwall-type energy inequality. We will derive such inequality by \emph{a priori} estimate method for simplicity. Let $(\eta,u)$ be a solution
of the RT problem, such that
\begin{equation}\label{aprpiosesnewxx}
 \sup_{0\leqslant t < T}\sqrt{\|(\varrho(t),u(t))\|_4^2}\leqslant \delta \in (0,1)\;\;\mbox{ for some  }T.
\end{equation}
Moreover, the solution enjoys fine regularity, which makes sure the procedure of formal deduction.
In addition, we rewrite \eqref{0105} with the boundary-value condition in \eqref{0107} as a nonhomogeneous  form:
\begin{equation}\label{0105xx}\left\{\begin{array}{l}
\varrho_t+   \bar{\rho}'u_3=\mathcal{N}^1:=-  u\cdot\nabla\varrho, \\[1mm]
 \bar{\rho} u_t+\nabla q-\mu\Delta{ {{u}}}+g\varrho {e}_3=\mathcal{N}^2:=-(\varrho+\bar{\rho})  u\cdot\nabla
 u-\varrho   u_t,\\[1mm]
 \mathrm{div} {u}=0,\\
u|_{\partial\Omega}=0.\end{array}\right.  \end{equation}

Next we establish zero estimate of $u$, space derivative estimate of $\varrho$,   temporal derivative estimate of $u$,
Stokes estimates of $(u,u_t)$, and equivalence estimate of $\mathcal{E}$ in sequence.
\begin{lem}\label{201612132242nnxx}
Under the assumption \eqref{aprpiosesnewxx}, it holds that
\begin{align}
&
\frac{\mm{d}}{\mm{d}t} \|\sqrt{\bar{\rho}}u\|_0^2  +
c\|u\|^2_1   \lesssim
\|(\varrho,  u_3)\|^2_{0}
 + \sqrt{\mathcal{E}} \mathcal{D}.
\label{ssebdaiseqinM0846x}   \end{align}
\end{lem}
\begin{pf}
 Multiplying  \eqref{0105xx}$_2$ by   $u$ in $L^2$ (i.e., taking inner product of \eqref{0105xx}$_2$ and   $u$ in $L^2$), then, using integration by parts,  we get
\begin{align}
 \frac{\mm{d}}{\mm{d}t} \|\sqrt{\bar{\rho}} u\|_0^2
 +\mu\|\nabla u\|_0^2=-g\int  \varrho u_3 \mm{d}x+ \int \mathcal{N}^2\cdot u\mm{d}x .
 \label{estimforhoedsds1stnxx}
\end{align}
By \eqref{aprpiosesnewxx}, it holds that
$$\int \mathcal{N}^2\cdot u\mm{d}x\lesssim \sqrt{\mathcal{E}}\mathcal{D}.$$
Thus we immediately derive \eqref{ssebdaiseqinM0846x} from \eqref{estimforhoedsds1stnxx} by using the above estimate,  and Young's and  Friedrichs's inequalities. \hfill$\Box$
\end{pf}
\begin{lem}\label{201612132242nnxxxx}
Under the assumption \eqref{aprpiosesnewxx}, it holds that
\begin{align}
&
\frac{\mm{d}}{\mm{d}t}\|\varrho\|_4^2  \lesssim
\|\varrho\|_4\|u\|_4.\label{ssebdaiseqinM0846xxx}   \end{align}
\end{lem}
\begin{pf}
Let $\alpha:=(\alpha_1,\alpha_2,\alpha_3)$ be a multiindex with order $|\alpha|:=\alpha_1+\alpha_2+\alpha_3\leqslant 4$, and  $\partial^{\alpha}:=\partial^{\alpha_1}_1\partial^{\alpha_1}_2\partial^{\alpha_1}_3$.

Noting that
$$\int  u \cdot\nabla  \partial^{\alpha}  \rho   \partial^{\alpha}\rho\mm{d}x=0,$$
thus, applying $\partial^{\alpha}$ to \eqref{0105xx}$_1$, and then multiplying the resulting identity by $\partial^{\alpha} \varrho$ in $L^2$, we get
\begin{align}
 \frac{\mm{d}}{\mm{d}t} \|\partial^{\alpha}\varrho\|_0^2
 =-\int \partial^{\alpha}(  \bar{\rho}' u_3 ) \partial^{\alpha}\varrho \mm{d}x+ \int (u \cdot\nabla  \partial^{\alpha}  \rho -\partial^{\alpha}(u\cdot\nabla \varrho))  \partial^{\alpha}\rho\mm{d}x=:I_{1,\alpha}.
 \label{estimforhoedsds1stnxxxx}
\end{align}
By \eqref{aprpiosesnewxx}, it holds that
$$I_{1,\alpha}\lesssim \|\varrho\|_4\|u\|_4\mbox{ for any }0\leqslant  |\alpha|\leqslant 4.$$
Thus we immediately derive \eqref{ssebdaiseqinM0846xxx} from \eqref{estimforhoedsds1stnxxxx}. \hfill$\Box$
\end{pf}

\begin{lem}\label{201612132242nxsfssdfsxx}
 Under the assumption \eqref{aprpiosesnewxx}, it holds that
\begin{align}
&   \frac{\mm{d}}{\mm{d}t} \|\sqrt{\varrho+\bar{\rho}}  u_t \|_{0}^2
+ c\|   u_t \|_{1}^2  \lesssim  \|  u_3\|_{ 0}^2+ \sqrt{\mathcal{E} } \mathcal{D},\label{Lem:030dsfafds1m0dfsf832xxxxx}\\
 &  \frac{\mm{d}}{\mm{d}t} \|\sqrt{\varrho+\bar{\rho}}   u_{tt} \|_{0}^2
+ c\|  u_{tt} \|_{1}^2  \lesssim  \| \partial_t u_3\|_{ 0}^2+ \sqrt{\mathcal{E} } \mathcal{D} .\label{Lem:030dsfafds1m0dfsf832xxx}
 \end{align}
\end{lem}
\begin{pf}
Let $1\leqslant i\leqslant 2$. Applying $\partial^i_t$ to \eqref{0105}$_2$, \eqref{0105}$_3$ and \eqref{0105xx}$_4$, we get
\begin{equation}\label{0105xxsdfs}\left\{\begin{array}{l}
\partial^i_t((\varrho+\bar{\rho}){  u}_t)+\partial^i_t((\varrho+\bar{\rho}){  u}\cdot\nabla
 u)+\nabla \partial^i_tq=\mu\Delta \partial^i_t u-g\partial^i_t\varrho {e}_3,\\
 \mathrm{div}\partial^i_t {u}=0 ,\\[1mm] \partial_t^iu|_{\partial\Omega}=0.\end{array}\right.  \end{equation}

Multiplying \eqref{0105xxsdfs}$_1$ with $i=2$ by $u_{tt}$ in $L^2$, and using the integration by parts and \eqref{0105xx}$_1$, we can compute out that
\begin{align}
&\frac{1}{2}\frac{\mm{d}}{\mm{d}t}\left(\|\sqrt{\varrho+\bar{\rho}} u_{tt}\|^2_0
 \right)+\mu\|\nabla u_{tt}\|_{0}^2\nonumber \\
&=\int (g \partial_t(\bar{\rho}'u_3-\mathcal{N}^1)e_3-  \partial_t^2 ((\varrho+\bar{\rho}){  u}\cdot\nabla
 u))  \cdot u_{tt}\mm{d}y\nonumber
 \\
 &\quad +\frac{1}{2}\int \varrho_tu_{tt}^2\mm{d}x+\int ((\varrho+\bar{\rho})u_{ttt}-\partial_{t}^2((\varrho+\bar{\rho}) u_t))\cdot u_{tt}\mm{d}x = :I_{2}. \label{0425sdfasfazz}
\end{align}

By \eqref{aprpiosesnewxx}, it holds that
$$|I_{2}|\lesssim \|   u_{tt}\|_0 (\| \partial_tu_3\|_0+ \sqrt{\mathcal{E}\mathcal{D}}).$$
Putting the above estimate into \eqref{0425sdfasfazz}, and then using  Friedrichs's and Young's inequalities, we get \eqref{Lem:030dsfafds1m0dfsf832xxx}.
Similarly, we can easily derive \eqref{Lem:030dsfafds1m0dfsf832xxxxx} from \eqref{0105xxsdfs} with $i=1$.
  \hfill$\Box$
 \end{pf}
\begin{lem}\label{201612132242nxxx}
Under the assumption \eqref{aprpiosesnewxx} with sufficiently small $\delta$,
\begin{align}
  &\|(u,q)\|_{\mm{S},2} \lesssim    \|\varrho\|_2+\|(u, u_{tt})\|_0,\label{2017020614181721xxxx}   \\
  &\|\partial_t(u,q)\|_{\mm{S},0} \lesssim
\|  u\|_0+(\|\varrho\|_2+\|u\|_0)\|u\|_1  +\| u_{tt}\|_0   ,\label{2017020614181721xxxxxx} \\
& \|(u,q)\|_{\mm{S},5} \lesssim  \| \varrho\|_3 + \| (u_3,u_{tt})\|_1+\sqrt{\mathcal{\mathcal{E}\mathcal{D}}}, \label{201901141547} \\
& \|\partial_t(u,q)\|_{\mm{S},3} \lesssim\| (u_3,u_{tt})\|_1+\sqrt{\mathcal{\mathcal{E}\mathcal{D}}}\label{2017020614181721xx1}.
   \end{align}
\end{lem}
\begin{pf}
Applying $ \partial_t^i$ to \eqref{0105xx}$_2$, we have
\begin{equation}\label{06051331xx}
\bar{\rho} \partial_t^{i+1} u +\nabla  \partial_t^{i} q -\mu\Delta  \partial_t^{i} u
=  \partial_t^{i}\mathcal{N}^2 -g \partial_t^{i}\varrho e_3.
 \end{equation}
By \eqref{0105xxsdfs}$_2$, \eqref{0105xxsdfs}$_3$ and  \eqref{06051331xx}, we get Stokes problem: for $i=0$ and $1$,
\begin{equation}\label{n0101nn928xx}\left\{\begin{array}{ll}
\nabla  \partial_t^{i}  q-\mu\Delta \partial_t^{i}  u= \mathcal{M}^1 , \\[1mm]
\mathrm{div} \partial_t^{i} u =0 ,\\
 \partial_t^{i} u|_{\partial\Omega}=0,
\end{array}\right.\end{equation}
where we have defined that
\begin{align}\label{201812271801}
&\mathcal{M}^1:=\left\{
                  \begin{array}{ll}
- g \varrho e_3-\bar{\rho}   u_t  + \mathcal{N}^2 & \hbox{ for }i=0, \\
               g  (\bar{\rho}'u_3-\mathcal{N}^1)e_3-\bar{\rho}  u_{tt}  + \mathcal{N}^2_t & \hbox{ for }i=1.
                  \end{array}
                \right.
\end{align}

Applying the classical Stokes estimate to \eqref{n0101nn928xx} yields that
\begin{align}
& \|  \partial_t^{i} (u,q)\|_{\mm{S},2-2i}  \lesssim \left\{
                  \begin{array}{ll}
\|(\varrho,  u_t,\mathcal{N}^2)\|_2 & \hbox{ for }i=0, \\
           \|( u_3,  u_{tt},\mathcal{N}^1, \mathcal{N}^2_t)\|_{0} & \hbox{ for }i=1
                  \end{array}
                \right.
\label{20171111841}
\end{align}
and
\begin{align}
& \|  \partial_t^{i} (u,q)\|_{\mm{S},3-2i}  \lesssim \left\{
                  \begin{array}{ll}
\|(\varrho,  u_t,\mathcal{N}^2)\|_3 & \hbox{ for }i=0, \\
           \|( u_3,  u_{tt},\mathcal{N}^1, \mathcal{N}^2_t)\|_{1} & \hbox{ for }i=1.
                  \end{array}
                \right.
\label{20171111841x}
\end{align}

By \eqref{aprpiosesnewxx},
we can estimate that
\begin{align}
&   \|(\mathcal{N}^1,\mathcal{N}_t^2)\|_0\lesssim (\|\varrho\|_2+\|\varrho_t\|_0)\|u\|_1+\|u_t\|_2(\|\varrho_t\|_0+\|u\|_1)+\|\varrho\|_2\|u_{tt}\|_0, \label{2018111959} \\
& \|\mathcal{N}^2 \|_{2}   \lesssim \|u\|_2\|u\|_3+\|\varrho\|_2\|u_t\|_2 , \label{201901101924}
\\
 & \|\mathcal{N}^2\|_3 +\|(\mathcal{N}^1,\mathcal{N}^2_t) \|_1 \lesssim    \sqrt{\mathcal{\mathcal{E}\mathcal{D}}}.\nonumber
\end{align}
In addition, using \eqref{aprpiosesnewxx} and \eqref{0105xxsdfs}$_1$, we have
\begin{equation}
\label{20181111212456}
\|\varrho_t\|_0\lesssim \|u\|_0.
\end{equation}
 Exploiting Young inequality and  the four estimates above, we get \eqref{2017020614181721xxxx}--\eqref{2017020614181721xx1} from \eqref{20171111841} and \eqref{20171111841x}.
\hfill $\Box$
\end{pf}
\begin{lem}
\label{lem:dfifessim2057sadfafdasxx}
Under the assumption \eqref{aprpiosesnewxx}  with sufficiently small $\delta$, we have
\begin{equation} \mathcal{E}\mbox{ is equivalent to }\|(\varrho,u)\|_4^2 \label{201702071610nbxx}
\end{equation}
\end{lem}
\begin{pf}
By \eqref{2017020614181721xxxx} and \eqref{2017020614181721xxxxxx}, to get \eqref{201702071610nbxx}, it suffices to derive, for sufficiently small $\delta$,
\begin{equation}
\label{201807311651xx}
\|\varrho_t\|_3+\|\varrho_{tt}\|_2 +\|u_{tt}\|_0\lesssim \|(\varrho,u)\|_4.
\end{equation}
Next we verify \eqref{201807311651xx}.

Multiplying \eqref{06051331xx}  with $i=1$ by $u_{tt}$ in $L^2$, we infer that
$$   \|\sqrt{\bar{\rho}} u_{tt}\|_0^2 =\int (g (\bar{\rho}'u_3-\mathcal{N}^1)e_3+\mu\Delta   u_t+  \mathcal{N}^2_t
 )\cdot u_{tt} \mathrm{d}y.
$$
 Using \eqref{0102},  \eqref{aprpiosesnewxx}, \eqref{2018111959}, \eqref{20181111212456} and Young's inequality, we can derive from the above identity that
\begin{equation} \label{uttestimsxxx}
\| u_{tt}\|_0^2 \lesssim \|u\|_1^2+\|u_{t}\|_2^2.
\end{equation}
Next we shall estimate for $\|u_t\|_2^2$.

Using \eqref{201901101924}, we can derive from \eqref{0105xx}$_2$ that
\begin{align}
\|u_t\|_2\lesssim \|\varrho\|_2+\|q\|_3+\|u\|_4,\label{201901012149}
\end{align}
and
\begin{equation}\label{n0101nn928xxxxx}\left\{\begin{array}{ll}
-\Delta q = \mathcal{M}^2:=\bar{\rho}'\partial_t u_3 +g\partial_3\varrho+\mm{div}((\varrho+\bar{\rho})  u\cdot\nabla
 u)+ u_t\cdot\nabla\varrho,  \\[1mm]
 \nabla q \cdot \vec{n}|_{\partial\Omega}=\mathcal{M}^3:= (\mu\Delta{ {{u}}}-g\varrho {e}_3)\cdot \vec{n},
\end{array}\right.\end{equation}
where $\vec{n}$ denotes the unit outer normal vector on $\partial\Omega$.
Apply the classical elliptic regularity theory to \eqref{n0101nn928xxxxx} yields
that
$$\|q\|_3\lesssim \|\mathcal{M}^2\|_1+\|\mathcal{M}^3\|_{H^{3/2}(\partial\Omega)}\lesssim \|(\varrho,u)\|_2+\|u_t\|_1.$$
Inserting the above estimate into \eqref{201901012149}, and then using interpolation inequality, we arrive at
\begin{align}
\|u_t\|_2\lesssim \|(\varrho,u)\|_4+\|u_t\|_0\label{201901012149xxxx}
\end{align}
Next we shall estimate for $u_t$.

We multiply \eqref{06051331xx}  with $i=0$ by $u_t$ in $L^2$, and then use the integration by parts to obtain
\begin{align}
\|\sqrt{ \bar{\rho} }u_{t}\|_{0}^2   = & \int  ( \mu\Delta{ {{u}}}+\mathcal{N}^2-g\varrho {e}_3 )\cdot u_t\mm{d}x \nonumber .\end{align}
 Using   \eqref{aprpiosesnewxx} and Young's inequality, we can derive from the above identity that
\begin{equation}   \label{utestimstst}
\| u_{t}\|_{0}^2\lesssim \|(\varrho, u )\|_2^2.
\end{equation}
Thus we derive from \eqref{uttestimsxxx}, \eqref{201901012149xxxx} and \eqref{utestimstst} that,  for sufficiently small $\delta$,
\begin{equation}
\label{201807311651xxsfa}
\|u_t\|_2+ \|u_{tt}\|_0\lesssim \|(\varrho,u)\|_4.
\end{equation}

Finally, by \eqref{0105xx}$_1$, we have
\begin{align}
&\|\varrho_t\|_3\lesssim \|u\|_3, \label{201901141546}  \\
& \|\varrho_{tt}\|_2\lesssim \|u_t\|_2 +\|u\|_2\|\varrho_t\|_3. \label{201901141546xx}
\end{align}
Thus we immediately get \eqref{201807311651xx} from  \eqref{201807311651xxsfa} and the two estimates above.
  \hfill$\Box$
\end{pf}

\begin{lem}
\label{201812271055}
For any given constant $\Lambda>0$, there exist  constants $\delta_1\in (0,1)$ and  $C_1>0$ such that, for any $\delta\leqslant \delta_1$,  then, under the assumption of \eqref{aprpiosesnewxx}, $(\varrho,u)$ satisfies the  Gronwall-type energy inequality,
\begin{equation}\label{energyinequality}
\mathcal{E}(t) +\frac{1}{C_1}\int_0^t {D}(\tau)\mm{d}\tau \leqslant
\Lambda \int_0^t\mathcal{E}(\tau)\mm{d}\tau +C_1\left(\|(\varrho^0,u^0)\|_4^2+\int_0^t\|(\varrho, {u})(s)\|_{0}^2\mm{d}s\right).
\end{equation}
\end{lem}
\begin{pf}
We can derive from Lemmas \ref{201612132242nnxx}--\ref{201612132242nxsfssdfsxx} that, for some sufficiently  large $c_1$,
\begin{align}
\label{20190102143}
\frac{\mm{d}}{\mm{d}t} \mathcal{E}_1+ c\mathcal{D}_1 \leqslant c_1( \|(\varrho,  u_3)\|^2_{0}
 +\|\varrho\|_4\|u\|_4+ \sqrt{\mathcal{E}} \mathcal{D})/c,
 \end{align}
 where we have defined that
 $$
\begin{aligned}
&\mathcal{E}_1:= \|\sqrt{\bar{\rho}}u\|_0^2+\|\sqrt{\varrho+\bar{\rho}}  (c_1 u_t,u_{tt})\|_{0}^2 +\|\varrho\|_4^2,\\
&\mathcal{D}_1:=  \| ( c_1u_t,u_{tt})\|_{1}^2+\| u\|^2_1 .
\end{aligned} $$

Using interpolation inequality and \eqref{201901141547},
\begin{align}
\|u\|_4\lesssim & \|u\|_0^{1/5}\|u\|_5^{4/5}\\
\lesssim & \|u\|_0^{1/5}(\| \varrho\|_3 + \| (u_3,u_{tt})\|_1+\sqrt{\mathcal{\mathcal{E}\mathcal{D}}})^{4/5}.
\end{align}
Integrating \eqref{20190102143} with respect to $t$, and then using Young's inequality, we get, for any given $\varepsilon>0$,
\begin{align}
&\mathcal{E}_1+c\int_0^t\left(\mathcal{D}_1+\|\varrho(\tau)\|_3^2\right)\mm{d}\tau\nonumber \\
&\leqslant \mathcal{E}_1|_{t=0}
+\varepsilon\int_0^t\|\varrho(\tau)\|_4^2\mm{d}\tau+\frac{1}{c}\int_0^t\left(\|(\varrho,  u)(\tau)\|^2_{0}  + \sqrt{\mathcal{E}} \mathcal{D}\right) \mm{d}\tau,
\label{201812291619}
\end{align}
where positive constant $c$ further depend on $\varepsilon$.

Noting that,  for sufficiently small $\delta$,
$$\|   (c_1 u_t,u_{tt})\|_{0}^2\lesssim  \|\sqrt{\varrho+\bar{\rho}}  (c_1 u_t,u_{tt})\|_{0}^2,$$
then, by \eqref{20181111212456}, \eqref{201901141546}, \eqref{201901141546xx} and Lemmas \ref{201612132242nxxx}--\ref{lem:dfifessim2057sadfafdasxx}, we easily observe that, for sufficiently small $\delta$,
\begin{align}
&\mathcal{E}_1, \ \mathcal{E}\mbox{ and }\|(\varrho,u)\|_4\mbox{ are equivalent}\mbox{ for any }t\geqslant 0, \label{201901110930} \\
& \mathcal{D}_1+\|\varrho \|_3^2 \mbox{ and }\mathcal{D}\mbox{ is equivalent}.\label{201901110930x}
\end{align}
Consequently we immediately derive \eqref{energyinequality} from \eqref{201812291619}--\eqref{201901110930x}.\hfill $\Box$
\end{pf}

Now we mention that the local existence of strong/classical solutions   to the 3D
nonhomogeneous incompressible Navier--Stokes equations have been
established, see \cite{CYKHOM1str,CYKHOM1ufda,TLYHCJ} for example. In
particular,  one can follow the proof  in \cite{TLYHCJ} with a slight modification to obtain a local existence result of a unique classical solution $(\varrho, {u})\in C^0([0,T^{\max}),H^4\times H^4_\sigma)$ to the RT problem \eqref{0105}--\eqref{0107} for some $T^{\max}>0$; moreover the classical solution satisfies \emph{ a priori} estimate in Lemma \ref{201812271055}.  The proof is standard by means of energy estimates, and hence we omit it here. Consequently,  we can arrive at the following conclusion:
\begin{pro} \label{pro:0401xx}
Let $\bar{\rho}\in C^5(\bar{\Omega})$, and  $K_1$, $K_2$ be two constants.
For any given initial data $(\varrho_0, {u}_0)\in H^4\times H^4_\sigma$ satisfying
\begin{align}
& \|(\varrho_0, {u}_0)\|_{4}\leqslant K_1 ,\\
&\label{201812291639} \inf_{ {x}\in\Omega}\{(\varrho_0+\bar{\rho})( {x})\}\geqslant K_2>0
\end{align}
 and  the compatibility condition
\begin{equation}
\label{201812291644}
\partial_t {u}^0|_{\partial\Omega}= {0},
 \end{equation}
 there exist a $T^{\max}>0$ (depending on $K_1$ and $K_2$) and a unique classical solution
$(\varrho, {u})\in C^0([0,T^{\max} ),H^4\times H^4_\sigma)$ to the RT
problem \eqref{0105}--\eqref{0107} with $q\in  C^0([0,T^{\max} ),\underline{H}^3)$. Moreover,
if $\mathcal{E}(t)\leqslant  {\delta}_1$ on $(0,T)\subset (0,T^{\max})$, then the classical solution satisfies \eqref{energyinequality}.
\end{pro}
\begin{rem}
For any given initial data $(\varrho_0, {u}_0)\in H^4\times H^4_\sigma$ satisfying  \eqref{201812291639}, there exists  a unique local-in-time strong solution $(\varrho, {u})\in C^0([0,T),H^2\times H^2_\sigma)$ with  a unique associated pressure $q\in C^0([0,T),\underline{H}^1)$. Moreover,  the initial date of the associated pressure $q$ is a weak solution to
\begin{equation}
\left\{
\begin{array}{ll}
\mm{div}\left(  \nabla q^0 /(\varrho^0+\bar{\rho})\right)=\mm{div}\left(
\left(\mu\Delta u^0-g\varrho^0 {e}_3\right)/(\varrho^0+\bar{\rho})-u^0\cdot\nabla
 u^0\right)&  \mbox{ in }\Omega,\\
\nabla q^0  \cdot \vec{n} = (\mu \Delta u^0-g\varrho^0 ) \cdot \vec{n}   &\mbox{ on }\partial\Omega.
\end{array}
\right. \label{201808040916x}
\end{equation}
If \eqref{201812291644} is further satisfied, i.e., $(\varrho^0,u^0,q^0)$ satisfies
\begin{equation}
\label{201812271841xxx}
\nabla q^0-\mu\Delta u^0+g\varrho^0 {e}_3=0\mbox{ on }\partial\Omega,
\end{equation}
then we can improve the regularity of $(\varrho,u,q)$ so that $(\varrho,u)$ is a classical solution defined on some time interval \cite{RSTEVIJAMSS}.
\end{rem}

\begin{rem}
\label{201811031411xx}
It should be noted that, for any classical solution $(\varrho,u,q)$ constructed by Proposition \ref{pro:0401xx}, and for any given $t_0\in (0,T^{\max})$,
$(\varrho,u,q)|_{t=t_0}$ automatically satisfies \eqref{201808040916x}--\eqref{201812271841xxx} with $( \varrho, u,q)|_{t=t_0}$ in place of $(\varrho^0, u^0,q^0)$. We take  $(\varrho, u,q)|_{t=t_0}$ as a new initial datum, then the new initial datum can also define a unique local-in-time classical solution $( \tilde{\varrho},\tilde{u},\tilde{q})$ constructed by Proposition
\ref{pro:0401xx}; moreover the initial date of $\tilde{q}$ is equal to $q|_{t=t_0}$ by unique solvability of \eqref{201808040916x}.
\end{rem}

\subsection{Construction of initial data for  nonlinear solutions} \label{201805302028}

Let us first recall the linear instability result of the RT problem  \eqref{0105}--\eqref{0107}.
\begin{pro}\label{thm:0101xx}
Under the assumptions of Theorem \ref{thm:0102xx}, zero solution  is unstable to  the linearized RT equations \begin{equation}\label{0108}
\left\{\begin{array}{ll}
 \varrho_t+ \bar{\rho}'u_3=0, \\[1mm]
  \bar{\rho} {u}_t +\nabla q=\mu\Delta{ {{u}}}-g\varrho {e}_3,\\[1mm]
  \mathrm{div} {u}=0.
\end{array}\right.\end{equation}  That is, there exists an unstable solution
$$({\varrho}, {u},{q}):=e^{\Lambda t}(-\bar{\rho}'\tilde{u}_3/\Lambda,\tilde{ {u}},\tilde{q})$$
 to \eqref{0108}, where $(\tilde{u},\tilde{q})\in H^4_\sigma \times \underline{H}^3$
 solves the following boundary problem
 \begin{equation}\label{0113}
\left\{
                              \begin{array}{ll}
\Lambda^2\bar{\rho}\tilde{u}+\Lambda\nabla \tilde{q}-\Lambda\mu\Delta\tilde{u}
=g
\bar{\rho}'\tilde{u}_3  {e}_3,\\[1mm]
\mathrm{div} \tilde{u}=0,\quad
\tilde{u}|_{\partial\Omega}=0
\end{array}
                            \right.
\end{equation} with a (largest) growth rate $\Lambda$ defined by
\begin{equation}\begin{aligned}
\label{0111nn} \Lambda^2=\sup_{w\in
{H}^1_\sigma}\frac{g \int \bar{\rho}' {{w}}_3^2\mm{d} {x}-\Lambda\mu
\int|\nabla
 { {w}}|^2\mm{d} {x}}{\int\bar{\rho}| { {w}}|^2\mathrm{d} {x}}
.
\end{aligned}\end{equation}
 Moreover, $\tilde{u}$ satisfies
 \begin{equation}
\label{201812271810}
\bar{\rho}'u_3\neq 0,\  \tilde{u}_{\mm{h}}\neq 0.
\mbox{ and }
\tilde{u}_3\neq  0
\end{equation}
\end{pro}
\begin{pf} If  $\Omega$ is a $C^2$-bounded domain, and
the density profile $\bar{\rho}\in C^1(\bar{\Omega})$ satisfies \eqref{0102} and \eqref{0103},  Jiang--Jiang have proved that
zero solution  is unstable to the linearized system, where
 $(\tilde{u},\tilde{q})\in H^2_\sigma \times \underline{H}^1$, see \cite[Proposition 2.3.]{JFJSO2014}. However, if $\bar{\rho}$ further belongs to $C^5(\bar{\Omega}) $, by the classical regularity theory on the Stokes equations (see \cite[TheoremIV.6.1]{galdi2011introduction}), we can easily see from \eqref{0113} that $(\tilde{u},\tilde{q})\in H^4_\sigma \times \underline{H}^3$, and thus  get Proposition \ref{thm:0101xx}. \hfill$\Box$
\end{pf}

For any given $\delta>0$,  let
\begin{equation}\label{0501xx}
\left(\varrho^\mm{a}, u^\mm{a} ,q^{\mm{a}}\right)=\delta e^{{\Lambda t}} (\tilde{\varrho}^0,\tilde{u}^0,\tilde{q}^0),
\end{equation}
where $(\tilde{\varrho}^0,\tilde{u}^0, \tilde{q}^0):=(-\bar{\rho}'\tilde{u}_3/\Lambda,\tilde{u},\tilde{q})$, and $(\tilde{u},\tilde{q})\in H^4_\sigma\times  \underline{H}^3$ comes from Proposition \ref{thm:0101xx}.
Then $\left(\varrho^\mm{a}, u^\mm{a},q^\mm{a}\right)$  is a solution to the linearized RT   equations, and enjoys the estimate, for any given $i\geqslant 0$,
 \begin{equation} \nonumber
\|\partial_t^i(u^{\mm{a}}, q^{\mm{a}})\|_{\mm{S},2}\leqslant c(i) \delta e^{\Lambda t}.
 \end{equation}Moreover, by \eqref{201812271810},
 \begin{align}
&
\|\tilde{\varrho}^0\|_{L^1}\|\tilde{u}_{\mm{h}}^0\|_{L^1} \|\tilde{u}_3^0\|_{L^1}
  > 0.\nonumber
\end{align}
Unfortunately, the initial data of linear  solution $\left(\varrho^\mm{a}, u^\mm{a}, q^\mm{a}\right)$ does not satisfy the necessary compatibility conditions   of RT problem in general. Therefore, next we modify  the initial data of the linear solutions, so that the obtained new initial data satisfy \eqref{201808040916x} and  \eqref{201812271841xxx}.
\begin{pro}
\label{lem:modfiedxx}
Let $(\tilde{\varrho}^0,\tilde{u}^0,\tilde{q}^0)$
be the same as in \eqref{0501xx}, then there exists a constant $\delta_2$, such that for any $\delta\in(0,\delta_2)$, there is a couple $(u^{\mm{r}},q^{\mm{r}})\in H^4_\sigma\cap \underline{H}^3$ enjoying the following properties:
\begin{enumerate}[\quad (1)]
  \item  The modified initial data
  \begin{equation}\label{mmmode04091215xx}
  ( \varrho_0^\delta, {u}_0^\delta,q_0^\delta): =\delta
   ( \tilde{\varrho}^0, \tilde{u}^0, \tilde{q}^0) + \delta^2 (0,   u^\mm{r},q^{\mm{r}})
\end{equation}
belongs to $H^4_0\times H^4_\sigma\times \underline{H}^3$, and satisfies
the compatibility conditions   \eqref{201808040916x}$_1$ and \eqref{201812271841xxx}  with $(u_0^\delta,q_0^\delta)$ in place of $(u^0$, $q^0)$.
\item Uniform estimate:
\begin{equation}
\label{201702091755xx}
 \sqrt{ \|(u^{\mm{r}},q^{\mm{r}})\|_{\mm{S},2}^2 }\leqslant {C}_2,
 \end{equation}
where the constant $C_2\geqslant 1$ depends on the domain, the density profile  and the known parameters, but is independent of $\delta$.
\end{enumerate}
\end{pro}
\begin{pf}
Recalling the construction of $( \tilde{u}^0,\tilde{q}^0 )$, we can see that
$( \tilde{u}^0,\tilde{q}^0)$ satisfies
\begin{equation}
\label{201705141510xx}
\left\{
\begin{array}{ll}
 \mm{div}\tilde{u}^0 =0   &\mbox{ in }\Omega,\\
 \Lambda^2\bar{\rho} \tilde{u}^0 + \Lambda\nabla \tilde{q}^0- \Lambda\mu\Delta\tilde{u}^0= g\bar{\rho}'\tilde{u}_3^0  & \mbox{ in }\Omega,\\
 (\tilde{\eta}^0,\tilde{u}^0) = {0} & \mbox{ on }\partial\Omega.
 \end{array}\right.
\end{equation}
If $( u^\mm{r},q^\mm{r})\in H^4\times \underline{H}^3$ satisfies, for any given $\delta$,
\begin{equation}
\label{201702061320xxxxx}
\left\{
\begin{array}{ll}
 \mm{div} {u}^{\mm{r}}  =0   &\mbox{ in }\Omega,\\
 \nabla q^{\mm{r}}-\mu\Delta u^{\mm{r}} & \\
 =\mathcal{M}^2(\Upsilon,u^{\mm{r}}):=(\varrho_0^\delta+\bar{\rho}) (\Upsilon-(\delta \tilde{u}^0+\delta^2 u^{\mm{r}} ) \cdot\nabla
 (\delta \tilde{u}^0+\delta^2 u^{\mm{r}} ))/\delta^2&   \mbox{ in }\Omega,\\
  \mm{div}\Upsilon=- \delta \Lambda \mm{div}( {\varrho}_0^\delta\tilde{u}^0 /(\varrho^\delta_0+\bar{\rho}))&   \mbox{ in }\Omega,\\
(\Upsilon, \tilde{u}^{\mm{r}} ) = {0} & \mbox{ on }\partial\Omega.
\end{array}\right.\end{equation}
 where $(\varrho^\delta,{u} ^\delta,q ^\delta)$ is given in the mode \eqref{mmmode04091215xx}, then, by \eqref{201705141510xx}, it is easy to check that $(\varrho^\delta,{u} ^\delta,q ^\delta)$  belongs to $H^4_0\times H^4_\sigma \times \underline{H}^3$, and satisfies  \eqref{201808040916x}$_1$ and \eqref{201812271841xxx}  with $(u^\delta,q ^\delta)$ in place of $( u^0 ,q^0 )$. Next we construct such $( u^\mm{r},q^\mm{r} )$ enjoying \eqref{201702061320xxxxx} for sufficiently small $\delta$.

Since $H^2\hookrightarrow L^\infty$, there exists  a constant $\delta_3>0$ such that
\begin{equation}
\label{201901141558}
\inf_{x\in \overline{\Omega}}\{w+\bar{\rho}\}\geqslant \inf_{x\in \overline{\Omega}}\{\bar{\rho}\}/2>0\mbox{ for any }w\mbox{ satisfying }\|w\|_2\leqslant \delta_3.
\end{equation}
 We consider a  Stokes problem:
\begin{equation}
\label{201702051905xfweaxx}
 \left\{\begin{array}{ll}
\nabla q-   \Delta  \Upsilon  =0, \
 \mm{div}\Upsilon=- \delta \Lambda \mm{div}( {\varrho}_0^\delta\tilde{u}^0 /(\varrho^\delta_0+\bar{\rho}))  &\mbox{ in }\Omega, \\
\Upsilon=0 &\mbox{ on }\partial\Omega.
\end{array}\right.\end{equation}
Then, by the existence theory of Stokes problem, there exist a unique strong solution $( \Upsilon ,q)\in H^2\times  \underline{H}^1$
 to \eqref{201702051905xfweaxx} for any $\delta\leqslant \delta_3$;  moreover,
 \begin{align}
&\|(\Upsilon,q)\|_{\mm{S},0}\lesssim  \| \delta \Lambda \mm{div}( {\varrho}_0^\delta\tilde{u}^0 /(\varrho^\delta_0+\bar{\rho})) \|_{1}\lesssim\delta^2.
\label{Ellipticestimate0839nnxxxxx}
\end{align}

With $\Upsilon$ in hand, we further construct $(u^{\mm{r}},q^{\mm{r}})$. To this purpose, we consider a Stokes problem: for any give $w\in H_\sigma^4$,
\begin{equation}
\label{201702061320xxxx}
\left\{
\begin{array}{ll}
 \nabla q -\mu\Delta u  =\mathcal{M}^2( \Upsilon,w) &   \mbox{ in }\Omega,\\
 \mm{div} {u}   =0   &\mbox{ in }\Omega,\\
 {u}  = {0} & \mbox{ on }\partial\Omega,
\end{array}\right.\end{equation}
where $ \Upsilon$ is provided by \eqref{201702051905xfweaxx}.

Using the existence theory of Stokes problem again, there exist a solution $(u,q)\in H^4_\sigma\times  \underline{H}^3$
 to \eqref{201702061320xxxx};  moreover, by \eqref{Ellipticestimate0839nnxxxxx},
 \begin{align}
&\|(u,q)\|_{\mm{S},2}\lesssim  \| \mathcal{M}^2( \Upsilon,w )\|_2\leqslant c_2(1+\delta^2\|w\|_4^2).
\label{Ellipticestimate0839nn}
\end{align}
Therefore, one can construct an approximate function sequence $\{(u^{n}, q^n)\}_{n=1}^\infty$, such that, for any $n\geqslant 2$,
\begin{equation}
\label{201702051905xxxx}
 \left\{\begin{array}{ll}
\nabla q^{n}-\mu \Delta u^{n}=\mathcal{M}^2( \Upsilon,w^{n-1}) ,\quad
\mm{div}u^{n}=0 &\mbox{ in }\Omega, \\
u^{n}=0 &\mbox{ on }\partial\Omega,
\end{array}\right.\end{equation}
where $\|(u^1,q^1)\|_{\mm{S},2} \leqslant c_2$.   Moreover,  by \eqref{Ellipticestimate0839nn}, one has
$$ \|(u^n,q^n)\|_{\mm{S},2}   \leqslant  c_2(1+\delta^2 \|(u^{n-1},q^{n-1})\|_{\mm{S},2}^2 )$$
for any $n\geqslant 2$, which implies that  \begin{equation}
 \label{20170dfsf2090854}
\|(u^n,q^n)\|_{\mm{S},2} \leqslant 2c_2
 \end{equation} for any $n$, and any $\delta\leqslant 1/2c_2$.

 Next we further prove that  $\{(u^n,  q^n )\}_{n=1}^\infty$ is a Cauchy sequence.
Noting that
$$
 \left\{\begin{array}{ll}
\nabla (q^{n+1}-q^n)-\mu \Delta (u^{n+1}-u^{n})=\mathcal{M}^2( \Upsilon,u^n)-\mathcal{M}^2( \Upsilon,u^{n-1}),&\mbox{ in }\Omega,
\\
\mm{div}(u^{n+1}-u^{n})
=0 &\mbox{ in }\Omega, \\
u^{n+1}-u^{n}=0 &\mbox{ on }\partial\Omega,
\end{array}\right.$$
thus we can use Stokes estimate again to get that
\begin{align}
\|(u^{n+1}-u^{n } , q^{n+1}-q^{n })\|_{\mm{S},2}  \lesssim  & \| \mathcal{M}^2( \Upsilon,u^{n}  ) -\mathcal{M}^2( \Upsilon,u^{n-1}  )\|_2 \nonumber \\
\leqslant & c_2(1+\delta^2\| u^{n+1}-u^{n } \|_{4}^2),
\label{Ellipticestimate0837df}
\end{align}
which presents that $\{ (u^{ n}, q^n )\}_{n=1}^\infty $ is a Cauchy sequence in $H^4_\sigma\times \underline{H}^3$  by choosing a sufficiently small $\delta$.
Consequently, we can easily get
a limit function $(u^{\mm{r}},q^{\mm{r}})\in H^4_\sigma \times \underline{H}^3$, which solves
\begin{equation}
\nonumber
 \left\{\begin{array}{ll}
\nabla q^{\mm{r}}-\mu \Delta u^{\mm{r}} =\mathcal{M}^2( \Upsilon,u^{\mm{r}}  ), \
\mm{div}u^{\mm{r}}=0 &\mbox{ in }\Omega, \\
u^{\mm{r}}=0 &\mbox{ on }\partial\Omega
\end{array}\right.\end{equation}
by \eqref{201702051905xxxx}. Moreover,  by \eqref{20170dfsf2090854},
$  \|(u^{\mm{r}},q^{\mm{r}})\|_{\mm{S},2}  \leqslant 2c_2$, which  yields  \eqref{201702091755xx}. \hfill $\Box$
\end{pf}

\subsection{Error estimates and existence of escape times}\label{201807111049}

Let
$$
\begin{aligned}
&C_3:=   \|(\tilde{\varrho}^0,\tilde{u}^0)\|_4 +C_2\geqslant 1,\\
 & \delta<\delta_0:= \min\{\delta_1,2C_3\delta_2,\delta_3\}/2 C_3<1,
 \end{aligned}$$
and $(\varrho_0^\delta,u_0^\delta)$ be constructed by Proposition \ref{lem:modfiedxx}.
Recalling \eqref{201702091755xx}, \eqref{201901141558} and the relation $\varrho_0^\delta=\delta\tilde{\varrho}^0$,  we have
\begin{align}
 \|(\varrho_0^\delta, u_0^\delta)\|_4\leqslant C_3\delta<2C_3\delta_0 <\delta_3,
\label{201809121553xx}
\end{align}
then
$$\inf_{x\in \overline{\Omega}}\{\varrho_0^\delta+\bar{\rho}\}>0 .$$
Consequently, by Proposition \ref{pro:0401xx}, for any given $\delta<\delta_0$, there exists a (nonlinear) solution $(\varrho, u,q )$ of the RT problem defined on a time interval $I_{T^{\max}}$ with initial value $(  \varrho_0^\delta ,{u}_0^\delta )$.

Let $\epsilon_0\in (0,1)$ be a constant, which will be defined in \eqref{definedxx}.
 We define
 \begin{align}\label{timesxx}
& T^\delta:={\Lambda}^{-1}\mm{ln}({\epsilon_0}/{\delta})>0,\quad\mbox{i.e.,}\;
 \delta e^{\Lambda T^\delta }=\epsilon_0,\\
&T^*:=\sup\left\{t\in I_{T^{\max}}\left|~\|(\varrho, u) (\tau)\|_4 \leqslant 2C_3\delta_0\mbox{ for any }\tau\in [0,t)\right.\right\},\nonumber\\
&  T^{**}:=\sup\left\{t\in I_{T^{\max}}\left|~\left\|(\varrho,u)(\tau)\right\|_{0}\leqslant 2 C_3\delta e^{\Lambda \tau}\mbox{ for any }\tau\in [0,t)\right\}.\right.\nonumber
 \end{align}
 thus $T^*>0$ by \eqref{201809121553xx} and Proposition \ref{pro:0401xx}.  Similarly, we also have  $T^{**}>0$.
Moreover, by Proposition \ref{pro:0401xx} and Remark \ref{201811031411xx},
we can easily see that
 \begin{align}\label{0502n1xx}
& {\|(\varrho, u) (T^*)\|_4  }=2C_3 \delta_0,\mbox{ if }T^*<\infty ,\\
\label{0502n111}  & \left\|(\varrho,u) (T^{**}) \right\|_0
=2 C_3\delta e^{\Lambda T^{**}},\mbox{ if }T^{**}<T^{\max}.
\end{align}

We denote ${T}^{\min}:= \min\{T^\delta ,T^*,T^{**}\}$. Since $\|(\varrho,u)\|_4\leqslant \delta_1 $ in $I_{T^{\min}}$,
then $(\varrho,u)$ satisfies \eqref{energyinequality} in $I_{T^{\min}}$.
In addition,
$$\|(\varrho,u)(t)\|_0\leqslant 2C_3 \delta e^{\Lambda t}\mbox{ in }I_{T^{\min}}.$$ Thus we can easily derive from \eqref{energyinequality} that, for all $t\in I_{T^{\min}}
 $,
\begin{align}
\mathcal{E}(t)
\leqslant c\delta^2e^{2\Lambda t}+\Lambda\int_0^t  \mathcal{E}(\tau) \mm{d}\tau.
\label{0503xx} \end{align}
Applying Gronwall's inequality to the above estimate, we deduce that
\begin{align}
\mathcal{E}(t)
  \leqslant (C_4\delta^2 e^{2\Lambda t})^2\leqslant (C_4 \varepsilon_0)^2\mbox{ in } I_{T^{\max}}. \label{201702092114xx}
\end{align}

In addition, we have the following error estimate between the (nonlinear) solution $(\varrho,u )$ and the linear solution $(\varrho^\mm{a},u^{\mm{a}})$ provided by \eqref{0501xx}.
\begin{lem}\label{lem:0401xx}
 Let $\varepsilon_0\leqslant \delta_3/C_4$. There exists a constant $C_5$ such that, for  any  $\delta\in (0,1)$ and any $t\in I_{T^{\max}}$,
\begin{align}
\label{ereroexx}
&  \|(\varrho^{\mm{d}},u^{\mm{d}})\|_{\mathfrak{X}}   \leqslant C_5\sqrt{\delta^3e^{3\Lambda t}},
\end{align}
where $(\varrho^{\mathrm{d}}, u^{\mathrm{d}}):=(\varrho, u)-(\varrho^\mm{a},u^{\mm{a}})$, $ \mathfrak{X}=L^{1}$ or $L^2$, and $C_5$ is independent  of  $T^{\min}$.
\end{lem}
\begin{pf} Please refer to \cite[Lemma 3.1]{AJC51269} for the proof.
We mention that the condition ``$\varepsilon_0\leqslant \delta_3/C_4$" makes sure that $\inf_{x\in \overline{\Omega}}\{\varrho+\bar{\rho}\}\geqslant \inf_{x\in \overline{\Omega}}\{\bar{\rho}\}/2>0$.
\hfill$\Box$
\end{pf}

We define $m_0:=\min \{ \|\tilde{\varrho}^0\|_{L^1} ,\|\tilde{u}^0_{\mm{h}}\|_{L^1},\|\tilde{u}^0_3\|_{L^1}\} >0$, and
 \begin{equation}\label{definedxx}
\epsilon_0:=\min\left\{
 \frac{C_3\delta_0}{  C_4},
\frac{C_3^2}{4 C_5^2},
\frac{m_0^2}{4C_5^2}  \right\}>0.
 \end{equation}
Consequently, we further have the  relation
\begin{equation}
\label{201702092227xx}
T^\delta =T^{\min}\neq T^*\mbox{ or }T^{**},
\end{equation}
which can be showed by contradiction as follows:

If $T^{\min}=T^*$, then   $T^*<\infty$. Noting that $\epsilon_0\leqslant C_3\delta_0/C_4$, thus we deduce from \eqref{201702092114xx} that
\begin{equation*}\begin{aligned}
 \|(\varrho,u) (T^*)\|_4\leqslant C_3\delta_0< 2 C_3 \delta_0,
 \end{aligned} \end{equation*}
which contradicts \eqref{0502n1xx}. Hence, $T^{\min}\neq T^*$.

If $T^{\min}=T^{**}$, then   $T^{**}<T^*\leqslant T^{\max}$.
 Noting that $\sqrt{\epsilon_0}\leqslant C_3/2C_5$, we can deduce from  \eqref{0501xx},
\eqref{timesxx} and \eqref{ereroexx}   that \begin{equation*}\begin{aligned}
\| (\varrho,u)  (T^{**})\|_{0}
&\leqslant \|(\varrho^\mm{a},u^\mm{a})(T^{**})\|_{0}+\|(\varrho^\mm{d}, u^\mm{d}) (T^{**})\|_{0}\leqslant  \delta e^{{\Lambda T^{**}}}(C_3+ C_5\sqrt{\delta e^{\Lambda T^{**}}})\\
&\leqslant  \delta e^{{\Lambda T^{**}}}(C_3+ C_5\sqrt{\epsilon_0})
\leqslant 3C_3 \delta e^{\Lambda T^{**}}/2<2C_3\delta e^{\Lambda T^{**}},
 \end{aligned} \end{equation*}
which contradicts \eqref{0502n111}. Hence, $T^{\min}\neq T^{**}$.
We immediately see that \eqref{201702092227xx} holds.
Moreover, by the relation \eqref{201702092227xx} and the definition of $T^{\min}$, we see that $T^\delta<T^*\leqslant T^{\max}$, and thus $(\varrho,u,q)\in C^0(\overline{I_T},H^4\times  H^4_\sigma\times \underline{H}^3)$ for some $T\in (T^\delta,T^{\max})$.

Noting that $\sqrt{\epsilon_0}\leqslant m_0/2C_5$ and \eqref{ereroexx} holds for $t=T^\delta$, thus, making use of \eqref{0501xx}, \eqref{ereroexx}  and \eqref{timesxx}, we can easily deduce the following instability relation:
$$
 \begin{aligned}
\|  \chi (T^\delta)
\|_{{L^1}}\geqslant &
\| \chi   (T^\delta) \|_{{L^1}}
-
\|  \chi (T^\delta)
- \chi^{\mm{a}} (T^\delta)
\|_{{L^1}} \\
>&   \delta e^{\Lambda T^\delta }( \| \tilde{\chi}^{0} \|_{L^1}- C_5\sqrt{\delta e^{\Lambda T^\delta }}) \geqslant m_0 \epsilon_0 /2,\\
\end{aligned}
$$
where $\chi=\varrho$, $u_{\mm{h}}$ or $u_3$.
This completes the proof of  Theorem \ref{thm:0102xx} by taking $\epsilon:= m_0\epsilon_0 /2$.


\section{RT Instability in MHD fluids}\label{201812281036}

This section is devoted to the proof of Theorem \ref{201806dsaff012301}.
We shall introduce some simplified notations throughout this section.
\begin{align}
&\Omega_0^h:=\mathbb{R}^2\times (0,h),\ \|\cdot \|_{i} :=\|\cdot \|_{H^{i}(\Omega_{\mm{p}})},\ \partial_{\mm{h}}^i:=\partial_{1}^{\alpha_1}\partial_{2}^{\alpha_2}\mbox{ for }\alpha_1+\alpha_2=i,\nonumber \\
&  \int:=\int_{\Omega_{\mm{p}}},\ \|\cdot\|_{i,j}^2:=\sum_{\alpha_1+\alpha_2=i} \|\partial^{\alpha_1}_1\partial^{\alpha_2}_2\cdot\|_{j}^2, \  \|\cdot\|^2_{\underline{i},j}:=\sum_{0\leqslant k\leqslant i}\|\cdot\|_{i,j}^2 ,\nonumber \\
& \|(u,q)\|_{\mm{S},k}:=\sqrt{\|u\|_{k+2}^2+\|   q\|_{k+1}^2 } ,\  E(w):= \|\partial_{\bar{M}}w\|^2_0-g\int \bar{\rho}' |w_3|^2\mm{d}x,\nonumber \\
& \mathfrak{E}(t):=\|\eta(t)\|_5^2+\|(u,q)(t)\|_{\mm{S},2}^2+\|\partial_t
(u,q)\|_{\mm{S},0}^2+\|u_{tt}\|_0^2 ,\nonumber
\\ &  \mathfrak{D} (t):= \|\eta(t)\|_5^2+\|(u,q)(t)\|_{\mm{S},3}^2+\|\partial_t(u,q)\|_{\mm{S},1}^2+\|u_{tt}\|_1^2,  \nonumber\\
 & {H}_{0,*}^{k,1}:=\{  \varpi\in H^k_0(\Omega_{\mm{p}})~|~
\det\nabla \phi(y)=1, \phi (y): \overline{\Omega_0^h} \mapsto  \overline{\Omega_0^h } \mbox{ is a homeomorphism mapping},\nonumber \\
 &\qquad \qquad  \phi  (y): \Omega_0^h\mapsto  \phi (\Omega_0^h) \mbox{ is a } C^{k-2}\mbox{-diffeomorphic  mappings, where }\phi:=\varpi+y\},\nonumber \\
 &\det\mbox{ denotes the determinant of matrix}.\nonumber
\end{align}

In addition, $C_j^{j-l}$ denotes the  number of $(j-l)$-combinations from a given set $S$ of $j$  elements. The function space $X(\Omega_{\mm{p}})$ is simplified by $X$ for $X=L^p$, $H^i$  and so on. $a\lesssim b$ means that $a\leqslant cb$ for some constant $c>0$, which may depend on the domain $\Omega$, the density profile, and physical parameters  $\mu $,  $g$ and  $\bar{M}$.

\subsection{Transformed MRT problem}

Now we define   the flow map $\zeta$ as the solution  to
\begin{equation}
\left\{
            \begin{array}{ll}
\partial_t \zeta(y,t)=v(\zeta(y,t),t)&\mbox{ in }\Omega_{\mm{p}},
\\
\zeta(y,0)=\zeta^0(y)&\mbox{ in }\Omega_{\mm{p}}.
                  \end{array}    \right.
\label{201811191443}
\end{equation}
We denote the Eulerian coordinates by $(x,t)$ with $x=\zeta(y,t)$,
whereas $(y,t) $ stand for the
Lagrangian coordinates. Then we further define the Lagrangian unknowns by
\begin{equation*}
(\tilde{\varrho}, u,\tilde{p},B)(y,t)=(\rho,v,p+\lambda |M|^2/2,M)(\zeta(y,t),t).
\end{equation*}
In Lagrangian coordinates the evolution equations for  $\tilde{\varrho}$, $u$, $\tilde{p}$ and $B$ read as
\begin{equation}\label{0116} \left\{
                              \begin{array}{ll}
\zeta_t=u, \\[1mm]
\tilde{\varrho}_t=0,\\
\tilde{\varrho} u_t+\nabla_{\ml{A}}\tilde{p}-\mu \Delta_{\ml{A}}u=
\lambda B\cdot \nabla_{\ml{A}} B-\tilde{\varrho} ge_3, \\[1mm]
B_t=B\cdot\nabla_{\ml{A}}u, \\[1mm]
\div_\ml{A}u=\div_\ml{A}B=0
\end{array}
                            \right.
\end{equation}
with initial--boundary value conditions
\begin{equation}
\label{201809232118}
(\zeta,\tilde{\varrho}, u,B)|_{t=0}=(\zeta^0,\tilde{\varrho}^0, u^0,B^0) \mbox{ and } (\zeta ,u)|_{ \partial\Omega_{\mm{p}}} =(y,0).
\end{equation}
Next we explain the notations in the above initial-boundary value problem.

 The matrix $\mathcal{A}:=(\ml{A}_{ij})_{3\times 3}$ is defined via
\begin{equation*}
\ml{A}^{\mm{T}}=(\nabla \zeta)^{-1}:=(\partial_j \zeta_i)^{-1}_{3\times 3},
\end{equation*} where the subscript $\mm{T}$ denotes the transposition, and  $\partial_{j}$ denote the partial derivative with respect to the $j$-th components of variables $y$. $\tilde{\mathcal{A}}:= \mathcal{A}-I$, and $I$ is the $3\times 3$ identity matrix.
The differential operator $\nabla_{\ml{A}}$ is defined by $$\nabla_{\ml{A}}w:=(\nabla_{\ml{A}}w_1,\nabla_{\ml{A}}w_2,\nabla_{\ml{A}}w_3)^{\mm{T}}
\;\mbox{ and }\;\nabla_{\ml{A}}w_i:=(\ml{A}_{1k}\partial_kw_i,
\ml{A}_{2k}\partial_kw_i,\ml{A}_{3k}\partial_kw_i)^{\mm{T}}$$
for vector function $w:=(w_1,w_2,w_3)$, and the differential operator $\mm{div}_\ml{A}$ is defined by
\begin{equation*}
\mm{div}_{\ml{A}}(f^1,f^2,f^3)=(\mm{div}_{\ml{A}}f^1,\mm{div}_{\ml{A}}f^2,
\mm{div}_{\ml{A}}f^3)^{\mm{T}}
\mbox{ and }\mm{div}_{\ml{A}}f^i:=\ml{A}_{lk}\partial_k f_{l}^i
\end{equation*}
for vector function $f^i:=(f_{1}^i,f_{2}^i,f_{3}^i)^{\mm{T}}$. It should be noted that we have used the Einstein convention of summation over repeated indices. In addition, we define   $\Delta_{\mathcal{A}}X:=\mm{div}_{\ml{A}}\nabla_{\ml{A}}X$.

If the initial data $(\zeta^0,\tilde{\varrho}^0,B^0)$ satisfies
 \begin{equation*}
\tilde{\varrho}^0=\bar{\rho}(y_3)\mbox{  and }B^0=\partial_{\bar{M}}\zeta^0\ \mbox{(or }\ml{A}^{\mm{T}}_0 B^0=\bar{M}),
\end{equation*}
 then the initial-boundary value problem \eqref{0116}--\eqref{201809232118}
can be rewritten  as
a so-called transformed  MRT problem:
\begin{equation}\left\{\begin{array}{ll}
\eta_t=u &\mbox{ in } \Omega_{\mm{p}},\\[1mm]
\bar{\rho}u_{t}+\nabla_{\ml{A}}\sigma-\mu\Delta_{\ml{A}}  u=\lambda \partial_{\bar{M}}^2 \eta+gG e_3&\mbox{ in } \Omega_{\mm{p}}, \\[1mm]
\mathrm{div}_{\mathcal{A}} {u} = 0& \mbox{ in }  \Omega_{\mm{p}},\\
(\eta, u)=0 &\mbox{ on }\partial\Omega_{\mm{p}},\\
(\eta,u)|_{t=0}=(\eta^0,u^0)  &\mbox{ in } \Omega_{\mm{p}}
\end{array}\right.\label{n0101nnxxs} \end{equation}
where $\sigma=\tilde{p}-\bar{p}(\zeta_3)$,  $\eta:=\zeta-y$,
 $(\tilde{\varrho},B) =(\bar{\rho}(y_3), \partial_{\bar{M}} \zeta)$ and $G:=\bar{\rho}(y_3+\eta_3(y,t))-\bar{\rho}(y_3)$.
Then we  get classical solutions of RT instability of the transformed MRT problem.
\begin{thm}\label{thmsdfa3}
Under the assumptions of Theorem \ref{201806dsaff012301}, a zero solution to the transformed MRT problem \eqref{n0101nnxxs} is unstable in the Hadamard sense, that is, there are positive constants $\Lambda$, $m_0$, $\epsilon$ and $\delta_0$,
and a vector functions $(\tilde{\eta}^0,\tilde{u}^0,\tilde{\sigma}^0,\eta^\mm{r},u^\mm{r},\sigma^\mm{r})\in H^5_\sigma\times  H^4_\sigma\times \underline{H}^3\times H^5_0\times H_0^4\times \underline{H}^3$,
such that for any $\delta\in (0,\delta_0)$ and the initial data
\begin{equation}
\label{201901092134}
(\eta^0, u^0, \sigma^0):=\delta(\tilde{\eta}^0,\tilde{u}^0, \tilde{\sigma}^0)  + \delta^2(\eta^{\mm{r}},u^\mm{r}, \sigma^{\mm{r}})\in H^{5,1}_{0,*}\times H^4_\sigma\times \underline{H}^3,
\end{equation}
there is a unique classical solution $(\eta,u,\sigma)\in C^0(\overline{I_T},H^{5,1}_{0,*}\times  H^4_0\times \underline{H}^3)$ with initial data $(\eta^0,u^0,\sigma^0)$ to the  transformed MRT problem  satisfying
\begin{align}
 \|\chi_{\mm{h}}(T^\delta)\|_{0},\ \|\chi_3(T^\delta)\|_{L^1} \geqslant  {\varepsilon},  \label{201901092145xx}
\end{align}
and
\begin{align}
\|\partial_{\bar{M}}\chi_{\mm{h}}(T^\delta)\|_{L^1},\ \|\partial_{\bar{M}}\chi_3(T^\delta)\|_{L^1}\geqslant  {\varepsilon}\mbox{ if }
\bar{M}_3\neq0 \label{201901092145}
\end{align}
for some escape time $T^\delta:=\frac{1}{\Lambda}\mm{ln}\frac{2\epsilon}{m_0\delta}\in I_T$, where $\chi=\eta$ or $u$. Moreover, the initial data $(\eta^0,u^0,\sigma^0)  \in H^{5,1}_{0,*}\times H^4_\sigma\times \underline{H}^3$ satisfies the compatibility  conditions
\begin{alignat}{2}
&\mm{div}_{\mathcal{A}^0}u^0=0& &\ \mbox{ in }\Omega_{\mm{p}}, \label{20180923221611xx123}  \\
& \mm{div}_{\mathcal{A}^0}\left( u^0 \cdot \nabla_{\mathcal{A}^0} u^0 +( \nabla_{\mathcal{A}^0}\sigma^0-\mu \Delta_{\mathcal{A}^0}u^0- \lambda\partial_{\bar{M}}^2\eta^0- g G^0 e_3)/\bar{\rho}\right)=0 & &\ \mbox{ in }\Omega_{\mm{p}}, \\
&\nabla_{\mathcal{A}^0}\sigma^0-\mu \Delta_{\mathcal{A}^0}u^0 -\lambda \partial_{\bar{M}}^2\eta^0 -g G^0 e_3=0 & &\ \mbox{ on }\partial \Omega_{\mm{p}}, \label{20180923221611xx1234}
\end{alignat}
where $G^0:=\bar{\rho}(y_3+\eta_3^0)-\bar{\rho}(y_3)$, and  $\mathcal{A}^0$  denotes the initial data of $\mathcal{A}$, and is defined by $\eta^0$.
\end{thm}
\begin{rem}
\label{201809121705}
It should be noted that the solution $(\eta,u,\sigma)$ in the above theorem enjoys the additional regularity:
\begin{align}
& (u_t, u_{tt}, \sigma_t  ) \in C^0(\overline{I_T}, H^2_0 \times L^2\times \underline{H}^1), \label{201811142106}\\
& (u, u_t, u_{tt}, \sigma, \sigma_t ) \in L^2(I_T, H^5_0\times H^3_0\times H^1_0\times \underline{H}^4\times \underline{H}^2). \label{201811142107}
\end{align}
\end{rem}

Next we can prove the existence of classical solutions of RT instability for the transformed MRT problem by following the proof frame of Theorem \ref{thm:0102xx} with finer mathematical analysis.

\subsection{Gronwall-type energy inequality of nonlinear solutions}

Similarly to the content in Section \ref{201805081731},
 this section is devoted to establishing Gronwall-type energy inequality for solutions of the transformed MRT problem.
Let $(\eta,u)$ be a solution
of the transformed MRT problem, such that
\begin{equation}\label{aprpiosesnew}
 \sup_{0\leqslant t < T}\sqrt{\|\eta(t)\|_5^2+\|u(t)\|_4^2}\leqslant \delta \in (0,1)\;\;\mbox{ for some  }T
\end{equation}
and the initial data $\eta^0$ satisfies $\eta^0\in H^{5,1}_{0,*}$. Next we give some preliminary estimates involving $\mathcal{A}$.
\begin{lem}\label{201805141072}
Under the assumption \eqref{aprpiosesnew}, we have
 for $0\leqslant i\leqslant 4$ and for $0\leqslant j\leqslant 1$,
\begin{align}
&  \label{06041533fwqg}
\|\tilde{\mathcal{A}}\|_{i}\lesssim \| \eta\|_{i+1},\\
& \|\mathcal{A}_t\|_i \lesssim \|   u\|_{i+1} , \label{06142100x}\\
&  \label{06142100}
\|\ml{A}_{tt}\|_j \lesssim  \| ( u,u_t)\|_{j+1},\\
& \|\mathcal{A}_{ttt}\|_0\lesssim \|(u_t,u_{tt})\|_1.
\end{align}
Moreover, for sufficiently small $\delta$, we have
 \begin{align}
 &\label{20160614fdsa1957}
\|w\|_1 \lesssim \|\nabla_{\mathcal{A}}w\|_0  \mbox{ for any }\; w\in H^1_0.
\end{align}
\end{lem}
\begin{pf}We can deduce from \eqref{n0101nnxxs}$_1$ and \eqref{n0101nnxxs}$_3$ that $\partial_t \det \nabla (\eta+y)=0 $, which implies $\det (\nabla \eta+I)=1$ due to $\det(\nabla \eta^0+I)=1$.
Thus, by the definition of $\mathcal{A}$, we see that
\begin{equation}
\label{06131422}
\mathcal{A}=(A^*_{ij})_{3\times 3},
\end{equation}
where $A^{*}_{ij}$ is the algebraic complement minor of $(i,j)$-th entry of matrix $(\partial_j \zeta_i)_{3\times 3}$ with $\zeta:=\eta+y$. By \eqref{n0101nnxxs}$_1$, \eqref{06131422} and  Friedrichs's inequality, we can easily derive the desired estimates \eqref{06041533fwqg}--\eqref{20160614fdsa1957}. \hfill $\Box$
\end{pf}
\begin{rem}
By \eqref{06131422}, we can check that
\begin{equation*}  \partial_kA^{*}_{ik}=0\mbox{ or } \partial_k\mathcal{A}_{ik}=0,  \end{equation*}
which implies some important relations
\begin{equation}\label{diverelation}
\mm{div}_{\tilde{\mathcal{A}}}u=\mm{div} (\tilde{\mathcal{A}}^{\mm{T}}u)
\end{equation}
and, for $0\leqslant i\leqslant 2$ and $0\leqslant j\leqslant 1$,
\begin{equation}
\mm{div}_{\partial_t^i\mathcal{A}}\partial_t^ju=\mm{div} (\partial_t^i\mathcal{A}^{\mm{T}}\partial_t^j u).\label{20178011456115}\end{equation}
\end{rem}

Since $\eta\in H^5_0$ and $\det (\nabla \eta+I)=1$, then
\begin{equation}
\label{2019901122315}
\eta\in H^{5,1}_{0,*}\mbox{ for sufficiently small  }\delta,
 \end{equation}
please refer to Lemma 4.2 and  Remark 4.1 in \cite{JFJSOMITN} for the derivation.
Hence $\bar{\rho}(y_3+\eta_3)$ makes sense. Similarly to \eqref{0105xx}, we shall also  rewrite \eqref{n0101nnxxs} as a nonhomogeneous  form:
 \begin{equation}\label{20171102237xx}  \left\{\begin{array}{ll}
\eta_t=u &\mbox{ in } \Omega_{\mm{p}},\\[1mm]
\bar{\rho} u_t+ \nabla \sigma-\mu \Delta u-\lambda\partial_{\bar{M}}^2\eta -g\bar{\rho}'\eta_3 e_3 = \mathcal{N}^3 &\mbox{ in }  \Omega_{\mm{p}},\\[1mm]
\mathrm{div} {u} =- \mathrm{div}_{\tilde{\mathcal{A}}} {u} & \mbox{ in }  \Omega_{\mm{p}},\\
 (\eta,u)=0 &\mbox{ on }\partial\Omega_{\mm{p}} ,
\end{array}\right.\end{equation}
where we have defined
$$  \begin{aligned}
\mathcal{N}^3:=& \mu (\mm{div} \nabla_{\tilde{\mathcal{A}} }u+\mm{div}_{\tilde{\ml{A}}}\nabla_{\mathcal{A}}u )- \nabla_{\tilde{\ml{A}} }\sigma+ge_3\int_0^{\eta_3(y,t)}
\int_0^\tau\bar{\rho}''(y_3+s)\mm{d}s\mm{d}\tau.
\end{aligned}
$$
 In addition, applying $ \partial_t^j$ to \eqref{n0101nnxxs}$_2$--\eqref{n0101nnxxs}$_4$, and  using \eqref{20178011456115} and \eqref{20171102237xx}$_1$, we have that
\begin{equation}\label{01s06p}
\begin{cases}
\bar{\rho}\partial_t^{j+1} u+\nabla_{\mathcal{A}} \partial_t^j \sigma-\mu\Delta_{\mathcal{A}} \partial_t^ju
=\lambda  \partial_t^j\partial_{\bar{M}}^2\eta+
 g\bar{\rho}'\partial_t^{j-1} u_3  e_3 +\mathcal{N}^{3+j},\\[1mm]
\mathrm{div}_{\mathcal{A}} \partial_t^j u  =\mathrm{div} \mathcal{N}^{5+j} ,\\
\partial_t^j(\eta, u)|_{\partial\Omega_{\mm{p}}}=0,
\end{cases}
\end{equation}
where $1\leqslant j\leqslant 2$,  $\mathcal{N}^{3+j}:=\mathcal{N}_G^j
 +\mu N^{j}_u+N^{j}_q$,
\begin{align}
&\mathcal{N}_G^j:=g\partial_t^{j-1}\left(u_3\int_0^{\eta_3}\bar{\rho}''(y_3+s)\mm{d}s \right)e_3,\ \mathcal{N}^{j}_u:= \sum_{0\leqslant  m<j,\ 0\leqslant  n\leqslant j} \partial_t^{j-m-n }\mathcal{A}_{il}\partial_{l}(\partial_t^{n}\mathcal{A}_{ik}\partial_t^{m}
\partial_ku),\nonumber\\
&\mathcal{N}^{j}_\sigma:=-\sum_{0\leqslant l< j}(\partial^{j-l}_t\mathcal{A}_{ik}\partial^l_t \partial_k \sigma)_{3\times 1}\mbox{ and } \mathcal{N}^{5+j}:= \left(-\sum_{0\leqslant l<j}C_j^{j-l}\partial_t^{j-l} \mathcal{A}_{ki}\partial_t^l\partial_i u_k\right)_{3\times 1}.
\nonumber \end{align}
Then we establish the following estimates for nonlinear terms.
 \begin{lem}
\label{201806291049}
Under the assumption \eqref{aprpiosesnew},
\begin{enumerate}
  \item[(1)] the estimate of $\div { {\eta}}$: for $0\leqslant i\leqslant 4$ and for $1\leqslant j\leqslant 2$,
\begin{align}
&\label{improtian1}
\|\div { {\eta}}\|_{i} \lesssim \left\{
                                  \begin{array}{ll}
                                 \|\eta\|_2 \|\eta\|_{i+1}  & \hbox{ for }0\leqslant i\leqslant 1; \\
                           \|\eta\|_{i} \|\eta\|_{i+1}   & \hbox{ for }2\leqslant i\leqslant 4,
                                  \end{array}
                                \right.
\\
& \|\div u\|_i\lesssim
 \left\{
                                  \begin{array}{ll}
                          \|\eta\|_3\|u\|_{i+1}   & \hbox{ for }0\leqslant i\leqslant 1; \\
                           \|\eta\|_{i+1} \|u\|_{i+1}   & \hbox{ for }2\leqslant i\leqslant 4,
                                  \end{array}
                                \right.
\label{201808181500}\\
& \|\div u_t\|_j\lesssim \|\eta\|_{3}\|u_t\|_{j+1}+\|u\|_3^2.\label{201808181500x}
\end{align}
 \item[(2)] for sufficiently small $\delta$, we have, for $0\leqslant i\leqslant 3$, and for $0\leqslant j \leqslant 1$,
\begin{align}
\label{06011733}
& ¡¡\|\mathcal{N}^3\|_i\lesssim  \|\eta\|_4(\|\eta\|_{i}+\|(u,\sigma)\|_{\mm{S},i} ) ,\\&
¡¡\|\mathcal{N}^3_t\|_{j}\lesssim   \|(\eta,u)\|_{3}(\|(u,\sigma)\|_{\mm{S},j}+\|\partial_t(u,\sigma)\|_{\mm{S},j} ) ,\label{06011733xx}\\
& \|\mathcal{N}^{4} \|_0+\|\mathcal{N}^{6} \|_0
\lesssim \|u\|_3(\|\eta\|_0+\|(u,q)\|_{\mm{S},0}) \label{201806291xx},\\
& \|\mathcal{N}^{5} \|_0+\|\mathcal{N}^7_t \|_0
\lesssim \sqrt{\mathcal{E}\mathcal{D}},\label{201806291}\\
& \|\mathcal{N}^7\|_0\lesssim \|u\|_3\|(u,u_t)\|_1. \label{201806291xxx}\end{align}
 \end{enumerate}
\end{lem}
\begin{pf}
Since $\det (\nabla \eta+I)=1$, we can check that
\begin{equation}
\label{201812091021}
 \Psi(\eta):=\left(\begin{array}{c}
         - \eta_1(\partial_2\eta_2+\partial_3\eta_3 )+
           \eta_1(\partial_2\eta_3 \partial_3\eta_2
- \partial_2\eta_2\partial_3\eta_3) \\
      \eta_1\partial_1\eta_2  - \eta_2\partial_3\eta_3+
        \eta_1( \partial_1\eta_2 \partial_3\eta_3
-\partial_1\eta_3\partial_3\eta_2)\\
          \eta_1\partial_1\eta_3
+\eta_2\partial_2\eta_3+\eta_1(\partial_1\eta_3 \partial_2\eta_2
-\partial_1\eta_2\partial_2\eta_3)
        \end{array}\right)\mbox{ and }\mm{div}\eta=\mm{div}\Psi(\eta).
\end{equation}
Thus we can easily derive \eqref{improtian1}--\eqref{201808181500} from \eqref{20171102237xx}$_1$, \eqref{20171102237xx}$_3$ and \eqref{201812091021}.
Exploiting   \eqref{06041533fwqg}--\eqref{06142100} and \eqref{20171102237xx}$_1$,
we can derive the estimates \eqref{06011733}--\eqref{201806291xxx} from
 the definitions of $\mathcal{N}^3$--$\mathcal{N}^7$.
 \hfill $\Box$\end{pf}

Next we further establish energy estimates of $(\eta,u)$, which include  the $y_{\mm{h}}$-derivative  estimates of $(\eta,u)$ (see Lemmas \ref{201612132242nn}), the estimate of temporal derivative of $u$ (see Lemma \ref{201612132242nxsfssdfs}), the Stokes estimates of $u$ (see Lemma \ref{201612132242nx}), the hybrid derivative  estimate of $(\eta,u)$ (see Lemma \ref{lem:dfifessim2057}), and the equivalent estimate of $\mathcal{E}$ (see Lemma \ref{lem:dfifessim2057sadfafdas}). Next we will establish these estimates  in sequence.
\begin{lem}\label{201612132242nn}
Under the assumption \eqref{aprpiosesnew} with sufficiently small $\delta$,
                        for $0\leqslant i\leqslant 4$,
\begin{align}
&
\frac{\mm{d}}{\mm{d}t}\int  \left( \bar{\rho}\partial_\mm{h}^i \eta \cdot  \partial_\mm{h}^i u  + \frac{ \mu  }{2}|\nabla \partial_\mm{h}^i \eta |^2\right) \mm{d}y+
 \lambda  \|\partial_{\bar{M}}\partial_{\mm{h}}^i \eta\|_0^2 \lesssim
\|( \eta,  u)\|^2_{i,0}
 + \sqrt{\mathfrak{E}} \mathfrak{D} ,
\label{ssebdaiseqinM0846} \\
&
 \label{201702061418}
 \frac{\mm{d}}{\mm{d}t}(\|\sqrt{\bar{\rho}} \partial_\mm{h}^i u\|^2_0
  +\|\partial_{\bar{M}}\partial_{\mm{h}}^i \eta\|_0^2)
+ c\|\partial_\mm{h}^i   u \|_{1}^2 \lesssim
\| \eta_3\|_{i,0}^2+ \sqrt{\mathfrak{E} }\mathfrak{D} . \end{align}
\end{lem}
\begin{pf}
 (1) Applying $\partial_{\mm{h}}^i$ to  \eqref{20171102237xx}$_2$ and \eqref{20171102237xx}$_4$,  we have
 \begin{equation}\label{20171102237}  \left\{\begin{array}{ll}
\bar{\rho}\partial_{\mm{h}}^i u_t =\partial_{\mm{h}}^i(\mu\Delta u- \nabla \sigma+g\bar{\rho}'\eta_3e_3+\partial_{\bar{M}}^2\eta +\mathcal{N}^3 )  &\mbox{ in }  \Omega_{\mm{p}},\\[1mm]
 \partial_{\mm{h}}^i(\eta,u)=0 &\mbox{ on }\partial\Omega_{\mm{p}} .
\end{array}\right.\end{equation}
Multiplying \eqref{20171102237}$_1$  by $\partial^i_\mm{h}\eta$ in $L^2$ yields that
\begin{align}
&\frac{\mm{d}}{\mm{d}t}\int\left( \bar{\rho} \partial_{\mm{h}}^i \eta \cdot \partial_{\mm{h}}^i u +\frac{ \mu }{2}|\nabla  \partial_{\mm{h}}^i \eta|^2\right) \mm{d}y +\lambda\|\partial_{\bar{M}}\partial_{\mm{h}}^i \eta\|_0^2\nonumber  \\
&=\int ( { \bar{\rho} } |\partial_{\mm{h}}^i u|^2
+g \bar{\rho}'|\partial_{\mm{h}}^i\eta_3|^2 )\mm{d}y
 +\int \partial_{\mm{h}}^i \sigma\mm{div}\partial_{\mm{h}}^i\eta\mm{d}y +\int  \partial_{\mm{h}}^i{\mathcal{N}}^3 \cdot  \partial_{\mm{h}}^i\eta\mm{d}y.\nonumber
 \end{align}
 By  \eqref{improtian1}, \eqref{06011733}  and the integration by parts, we can easily get \eqref{ssebdaiseqinM0846}.

(2)
Multiplying \eqref{20171102237}$_1$  by $\partial^i_\mm{h}u$ in $L^2$ yields that
\begin{align}
&\frac{1}{2}\frac{\mm{d}}{\mm{d}t} \left( \|\sqrt{\bar{\rho}} \partial_{\mm{h}}^i u\|_0^2 +\lambda\|\partial_{\bar{M}}\partial_{\mm{h}}^i \eta\|_0^2\right)+ { \mu } \|\nabla  \partial_{\mm{h}}^i u\|^2_0\nonumber  \\
&=\int
 \partial_{\mm{h}}^i(g\bar{\rho}'\eta_3e_3  +\mathcal{N}^3 )  \cdot \partial^i_\mm{h}u \mm{d}y
 + \int \partial_{\mm{h}}^i \sigma\mm{div}\partial_{\mm{h}}^i u\mm{d}y .\nonumber
\end{align}
Thus, following the argument of \eqref{ssebdaiseqinM0846}, we can easily get \eqref{201702061418}  by further using \eqref{201808181500}, \eqref{06011733} and the integration by parts.
\hfill$\Box$
 \end{pf}
\begin{lem}\label{201612132242nxsfssdfs}Under the assumption \eqref{aprpiosesnew} with sufficiently small $\delta$,
\begin{align}
&  \frac{\mm{d}}{\mm{d}t}( \|\sqrt{\bar{\rho}}  u_{t}\|_{0}^2 +\|\partial_{\bar{M}}u\|_0^2 )
+ c\| u_t\|_{1}^2 \lesssim  \|  u_3\|_0^2+ \sqrt{\mathfrak{E} } \mathfrak{D} , \label{Lem:030dsfafds1m0dfsf832}\\
&  \frac{\mm{d}}{\mm{d}t}\left(\|\sqrt{\bar{\rho}}  u_{tt}\|_{0}^2+
\|\partial_{\bar{M}}u_t\|_0^2
+\int \nabla \sigma_t\cdot \mathcal{N}^7\mm{d}y
\right)
+ c\| u_{tt}\|_{1}^2 \lesssim  \|
 \partial_t u_3\|_{ 0}^2+ \sqrt{\mathfrak{E} } \mathfrak{D} . \label{Lem:030dsfafds1m0dfsf832xx}
 \end{align}
\end{lem}
\begin{pf}
Multiplying \eqref{01s06p}$_1$ with $j=2$ by $  u_{tt}$ in $L^2$, and using the integration by parts, we have
\begin{align}
&\frac{1}{2}\frac{\mm{d}}{\mm{d}t}\left(\|\sqrt{\bar{\rho}} u_{tt}\|^2_0+\lambda\|\partial_{\bar{M}}u_{t}\|_0^2 \right)+ {\mu} \|\nabla_{\mathcal{A}} u_{tt}\|_{0}^2\nonumber \\
&=\int \left(  g\bar{\rho}'\partial_t u_3 e_3 +\mathcal{N}^{5} \right)\cdot u_{tt}\mm{d}y-
\int \nabla \sigma_{tt}  \cdot \mathcal{N}^{7}\mm{d}y. \label{0425sdfasfa}
\end{align}

Noting that
\begin{align}\nonumber
\int \nabla \sigma_{tt}  \cdot \mathcal{N}^{7} \mm{d}y=&
\frac{\mm{d}}{\mm{d}t} \int \nabla \sigma_t\cdot \mathcal{N}^7\mm{d}y
  +\int  \nabla \sigma_t  \cdot \mathcal{N}^7_t  \mm{d}y,
\end{align}
thus, by \eqref{20160614fdsa1957}, \eqref{201806291} and Young's inequality, we get  \eqref{Lem:030dsfafds1m0dfsf832xx} from \eqref{0425sdfasfa}.

Similarly, we can easily derive \eqref{Lem:030dsfafds1m0dfsf832} from \eqref{01s06p}$_1$ with $j=1$ by using \eqref{01s06p}$_2$ and \eqref{201806291xx}.
  \hfill$\Box$
 \end{pf}
\begin{lem}\label{201612132242nx}
Under the assumption \eqref{aprpiosesnew} with sufficiently small $\delta$,
\begin{enumerate}
  \item[(1)] the following estimates hold:
\begin{align}
  &\|(u,\sigma)\|_{\mm{S},2}+\|\partial_t(u,\sigma)\|_{\mm{S},0} \nonumber \\
  &\lesssim|\bar{M}_3|   (\|\eta\|_4+\|u\|_0)+\|\eta\|_{\underline{2},2}+\|u\|_{\underline{2},0}+  \|u_{tt}\|_0+\|\eta\|_2 \|\eta\|_4,\label{2017020614181721}\\
&  \left\|  \left( u,\sigma\right)\right\|_{\mm{S},3}  \lesssim  |\bar{M}_3|  \|\eta\|_{5}   + \|\eta\|_{\underline{2},3} +\|u_t\|_3+\|\eta\|_3\|\eta\|_4,\label{2017020614181721xx}\\
& \left\| \partial_t \left( u,\sigma\right)\right\|_{\mm{S},1}  \lesssim
 |\bar{M}_3| \|u\|_3+ \|u\|_{\underline{2},1}+\|u_{tt}\|_1+\|(\eta,u)\|_3\|(u,\sigma)\|_{\mm{S},1}.
\label{2017020614181721xxx}
\end{align}
  \item[(2)]
if $\bar{M}_3=0$, we have
\begin{align}
 &\sum_{0\leqslant {\alpha_1}+{\alpha_2}\leqslant 2}\|\partial_1^{\alpha_1}\partial_2^{\alpha_2} (u,\sigma)\|_{\mm{S},1} \lesssim  \|\eta\|_{\underline{4},1} +\|u\|_{\underline{3},0}+\| u_{tt} \|_1+(\|\eta\|_5+\|u\|_3)\sqrt{\mathcal{D}},  \label{201901031537} \\
& \sum_{0\leqslant {\alpha_1}+{\alpha_2} \leqslant 1 } \|\partial_1^{\alpha_1}\partial_2^{\alpha_2}\partial_t(u,\sigma)\|_{\mm{S},0} \lesssim  \|u\|_{\underline{3},0} +\|u_{tt}\|_{1}+\|(\eta,u)\|_3\sqrt{\mathcal{D} } .\label{201901031537x} \end{align}
\end{enumerate}
\end{lem}
\begin{pf}
We rewrite \eqref{20171102237xx}$_2$--\eqref{20171102237xx}$_4$  as a  Stokes problem:
\begin{equation}\label{n0101nn928}\left\{\begin{array}{ll}
\nabla \sigma-\mu\Delta    u= \mathcal{M}^3 , \ \mathrm{div}  u =\mathrm{div} u & \mbox{ in }  \Omega_{\mm{p}},\\
 u=0 &\mbox{ on }\partial\Omega_{\mm{p}}  ,
\end{array}\right.\end{equation}
where we have defined that
\begin{align}
&\mathcal{M}^3:= \lambda\partial_{\bar{M}}^2\eta+ g\bar{\rho}'\eta_3 e_3-\bar{\rho} u_t +\mathcal{N}^3.
\end{align}

Applying  $\partial_t^j\partial_{\mm{h}}^k$  to \eqref{n0101nn928}, and then using  Stokes estimate, we have
\begin{align}
 \|  \partial_t^j\partial_{\mm{h}}^k(u,\sigma)\|_{\mm{S},i-2j-k}  \lesssim&
 |\bar{M}_3|\| \partial_t^j\partial_{\mm{h}}^k\eta\|_{i-2j-k+2} +  \|\partial_t^j\partial_{\mm{h}}^k\eta\|_{\underline{2},i-2j-k}\nonumber \\
 & + \| \partial_t^{j+1}\partial_{\mm{h}}^k u\|_{i-2j-k }+
  \|\partial_t^j  \partial_{\mm{h}}^k\mathcal{N}^3\|_{i-2j-k}+\| \partial_t^j \partial_{\mm{h}}^k  \mathrm{div} {u}\|_{i-2j-k+1} . \label{201801031521}
\end{align}

Taking $(i,j,k)=(2,0,0)$ and $(2,1,0)$ in \eqref{201801031521}, resp., we have
\begin{align}
& \|  (u,\sigma)\|_{\mm{S},2}  \lesssim
 |\bar{M}_3|\|  \eta\|_4 +  \| \eta\|_{\underline{2},2}  + \| u_{t}\|_{2}+
  \|  \mathcal{N}^3\|_2+\| \mathrm{div} {u}\|_{3}, \label{201801031521x}\\
&  \|\partial_t (u,\sigma)\|_{\mm{S},0}  \lesssim
 |\bar{M}_3|\|u\|_2 +  \| u\|_{\underline{2},0}  + \| u_{tt}\|_{0}+
  \| \mathcal{N}^3_t\|_0+\| \mathrm{div} {u}_t\|_{1} . \label{201801031521xx}\end{align}
Using \eqref{201808181500}--\eqref{06011733xx} and interpolation inequality, we derive \eqref{2017020614181721} from the two estimates above.

Similarly,  we can get \eqref{2017020614181721xx} and \eqref{2017020614181721xxx} from \eqref{201801031521} with $(i,j,k)=(3,0,0)$ and $(3,1,0)$, resp.

Finally, using \eqref{201808181500}--\eqref{06011733xx} again, we deduce \eqref{201901031537}--\eqref{201901031537x} from \eqref{201801031521} with $(i,j,k)=(3,0,2)$ and $(3,1,1)$. \hfill $\Box$
\end{pf}
\begin{lem}
\label{lem:dfifessim2057}
Under  the assumptions $\bar{M}_3\ne  0 $ and \eqref{aprpiosesnew} with sufficiently small $\delta$, there exists a functional  $\overline{\|\eta\|}_5$, which is equivalent to $\|\eta\|_5$ and satisfies
 \begin{align}
& \frac{\mm{d}}{\mm{d}t}
\overline{\|\eta\|}_5^2+
\left\|  \left( \eta,u\right)\right\|_5^2+
\|  \sigma \|_4^2 \lesssim  \| \eta \|_{\underline{4},1}^2+\|u\|_0^2+\| u_{tt} \|_1^2 .\label{201812301849} \end{align}
\end{lem}
\begin{pf}
We shall rewrite \eqref{n0101nn928}  as the following Stokes problem:
\begin{equation}\label{n0101nn928n}\left\{\begin{array}{ll}
\nabla \sigma -\mu \Delta \omega = \mathcal{M}^3- \lambda\bar{M}_3^2 \eta /\mu , \ \mathrm{div} \omega  = \mathrm{div}(u +\lambda \bar{M}_3^2 \eta /\mu) & \mbox{ in }  \Omega_{\mm{p}}  ,\\
 \omega =0 &\mbox{ on }\partial\Omega_{\mm{p}},
\end{array}\right.\end{equation}
where $\omega:= u +\lambda \bar{M}_3^2 \eta /\mu $.

Let  $0\leqslant j\leqslant 3$.
Applying $\partial_{\mm{h}}^j $ to \eqref{n0101nn928n} yields that
\begin{equation*}\label{n0101nn928nadfa}\left\{\begin{array}{ll}
\nabla \partial_{\mm{h}}^j   \sigma -\mu \Delta \partial_{\mm{h}}^j \omega = \partial_{\mm{h}}^j(  \mathcal{M}^3- \lambda\bar{M}_3^2 \eta /\mu) &\mbox{ in }  \Omega_{\mm{p}} , \\[1mm]
\mathrm{div}\partial_{\mm{h}}^j   \omega  =\partial_{\mm{h}}^j( \mathrm{div}(u +\lambda \bar{M}_3^2 \eta /\mu)) & \mbox{ in }  \Omega_{\mm{p}}  ,\\
 \partial_{\mm{h}}^j  \omega =0 &\mbox{ on }\partial\Omega_{\mm{p}} ,
\end{array}\right.\end{equation*}
Then
\begin{align}
 \| \omega \|_{j,5-j }^2+
\|    \sigma \|_{j,4-j }^2  \lesssim
\|    \mathcal{M}^3- \lambda\bar{M}_3^2 \eta \|_{j,3-j }^2+\|   \mathrm{div}(u +\lambda \bar{M}_3^2 \eta /\mu) \|_{j,4-j }^2.
 \label{20171111029}
 \end{align}
By \eqref{20171102237xx}$_1$, we further have
  \begin{align}
& \frac{\mm{d}}{\mm{d}t}
\left\| \sqrt{\lambda \bar{M}_3^2 /\mu}  \eta\right\|_{j,5-j}^2+\|\lambda \bar{M}_3^2   \eta/\mu \|_{j,5-j}^2+
\|    u\|_{j,5-j}^2+
\| \sigma\|_{j,4-j}^2\nonumber\\
&\lesssim
\|  \eta \|_{j+1,4-j}^2+\|u_{t}\|_{3}^2 + \|(\mm{div}\eta,\mm{div}u)\|_4^2 +\|\mathcal{N}^3\|_3^2.\label{20180627sdafa1120safaf}
 \end{align}
Making use of  \eqref{improtian1}, \eqref{201808181500}, \eqref{06011733}, \eqref{2017020614181721xxx}   and interpolation inequality,  we immediately derive \eqref{201812301849} from \eqref{20180627sdafa1120safaf}.  \hfill$\Box$
\end{pf}

\begin{lem}
\label{lem:dfifessim2057sadfafdas}
Under the assumption \eqref{aprpiosesnew} with sufficiently small $\delta$, we have
\begin{equation} \mathfrak{E}\mbox{ is equivalent to }\|\eta\|_5^2+\|u\|_4^2 \label{201702071610nb}
\end{equation}
\end{lem}
\begin{pf} Following the argument of \eqref{201702071610nbxx}, we can obtain \eqref{201702071610nb} from \eqref{20171102237xx}. The readers can also refer to the proof of \cite[Lemma 3.6]{JFJSJMFMOSERT}.
\hfill$\Box$
\end{pf}

Now we are in the position to derive  \emph{a prior} Gronwall-type energy inequality.
\begin{lem}  \label{pro:0301n0845}
For any given constant $\Lambda>0$, there exist  constants ${\delta}_4\in (0,1)$, $c>0$ and ${C}_6>0$ such that, for any $\delta\leqslant {\delta}_4$,  then, under the assumption of \eqref{aprpiosesnew}, $(\eta,u)$ satisfy the  Gronwall-type energy inequality,
for any $t\in I_T$,
\begin{align}
\tilde{\mathfrak{E}}(t)   \leqslant \Lambda\int_0^t
\tilde{\mathfrak{E}}(\tau) \mm{d}\tau +  {C}_6\left(\|\eta^0\|_5^2+ \|u^0\|_4^2 +\int_0^t
\|(\eta,u)\|_{0}^2\mm{d}\tau\right)
\label{2016121521430850}
\end{align}and the equivalent estimate
\begin{align}
\label{2018008121027}
\mathfrak{E}(t)\leqslant C_6\tilde{\mathfrak{E}}(t)\lesssim \mathfrak{E}(t).
\end{align}
\end{lem}
\begin{pf}
We can derive from Lemmas \ref{201612132242nn}--\ref{201612132242nxsfssdfs} that, for any sufficiently large constant  $c_3$,
 \begin{align}
 \label{201702061553}
\frac{\mm{d}}{\mm{d}t} \mathfrak{E}_1+c \mathfrak{O}_1\leqslant c_3  \left( \|\eta \|_{\underline{4},0}^2+\|u_3\|_0^2+
 \sqrt{{\mathfrak{E}}} {\mathfrak{D}} \right) /c,
 \end{align}
  where we have defined that $
\mathfrak{O}_1:= c_3( \|u_t\|_{1}^2+    \| u \|_{\underline{4},1}^2)+\|u_{tt}\|_{1}^2$ and
 $$\begin{aligned}\mathfrak{E}_1:=&\sum_{|\alpha|\leqslant 4}
  \bar{\rho}\partial_\mm{h}^{\alpha} \eta \cdot  \partial_\mm{h}^{\alpha} u    +
\frac{\mu}{2}\|\nabla \eta\|_{\underline{4},0}
+c_3(\|\sqrt{\bar{\rho}} u\|^2_{4,0}
  +\|\partial_{\bar{M}}  \eta\|_{4,0}^2  +\|\sqrt{\bar{\rho}}  u_{t}\|_{0}^2 +\|\partial_{\bar{M}}u\|_0^2) \\
 &+ \|\sqrt{\bar{\rho}}  u_{tt}\|_{0}^2+
\|\partial_{\bar{M}}u_t\|_0^2
+\int \nabla \sigma_t\cdot \mathcal{N}^7\mm{d}y.
\end{aligned}$$

(1) If $\bar{M}_3\neq0$,
we can further derive from \eqref{201812301849} and \eqref{201702061553} that
 \begin{align}
 \label{201702061553xx}
\frac{\mm{d}}{\mm{d}t} \mathfrak{E}_2+c \mathfrak{O}_2 \leqslant \frac{1}{c}\left(c_3^2 \left( \|\eta\|_{\underline{4},0}^2+\|u\|_0^2 +
 \sqrt{{\mathfrak{E}}} {\mathfrak{D}} \right)+ \|\eta\|_{\underline{4},1}^2\right)  ,
 \end{align}
where  $ \mathfrak{E}_2£º=c_3\mathfrak{E}_1+\overline{\|\eta\|}_5^2$ and $\mathfrak{O}_2:=c_3\mathfrak{O}_1+ \left\|  \left( \eta,u\right)\right\|_5^2+\|  \sigma \|_4^2$.

Making use of \eqref{201806291xxx}, \eqref{2017020614181721}, \eqref{2017020614181721xx}, \eqref{201702071610nb},   and Friedrichs's and Young's inequalities,
  there exists a constant $c$  such that,  for any  sufficiently large $c_3$ and for any  sufficiently small  $\delta_4$,
\begin{align}
& \mathfrak{E}_2,\ \mathfrak{E}\mbox{ and } \|\eta\|_5^2+\|u\|_4^2 \mbox{ are equivalent for any }t\geqslant 0,\\
\label{201808safdas02asdfsadf1219xx}
&c c_3  \|\eta\|_{\underline{4},1}^2  \leqslant  \mathfrak{E}_2,\\
\label{201808061629xx}&
\mathfrak{O}_2\mbox{ is equivalent to }\mathfrak{D}.
\end{align}
 Using interpolation inequality, we further deduce   the desired conclusions  \eqref{2016121521430850} and \eqref{2018008121027} from \eqref{201702061553xx}--\eqref{201808061629xx}.

(2)
Let $0\leqslant j\leqslant 3$ and $0\leqslant k\leqslant 3-j$.
Applying $\partial_{\mm{h}}^{k}\partial_3^{j}$ to \eqref{20171102237xx}$_2$, and then multiplying the resulting identity by $ \partial_{\mm{h}}^{k}\partial_3^{2+j}\eta$ in $L^2$,
we have
$$
 \frac{\mm{d}}{\mm{d}t}\|\sqrt{\mu } \partial_{\mm{h}}^k\partial_3^{2+j} \eta \|_{0}^2 = 2 \int \partial_{\mm{h}}^{k}\partial_3^j (\mu\Delta_{\mm{h}}u +\lambda \partial_{\bar{M}}^2\eta +g\bar{\rho}'\eta_3e_3 -\bar{\rho} u_t  -\nabla \sigma+ \mathcal{N}^3)\cdot \partial_{\mm{h}}^{k}\partial_3^{2+j}\eta \mm{d}y.
$$
Using \eqref{06011733}, we can estimate that
\begin{align}
\label{201901031937}
 \frac{\mm{d}}{\mm{d}t}\|\sqrt{\mu } \partial_3^{2+j} \eta \|_{\underline{3-j},0}^2  \lesssim &
 \| \sqrt{\mu}\partial_3^{2+j} \eta \|_{\underline{3-j},0}  \left(
\| \partial_3^j(\eta,u)\|_{\underline{5-j},0}+ \|\partial_3^j(\nabla \sigma,u_t)\|_{\underline{3-j},0} + \sqrt{ \mathfrak{E} \mathfrak{D}}\right) .
\end{align}

Exploiting Young inequality, \eqref{201901031537}  and \eqref{201901031537x}, we derive from the above estimate with $0\leqslant j\leqslant 1$ that
\begin{align}
& \frac{\mm{d}}{\mm{d}t} ( {\|\sqrt{\mu}\partial_3^2\eta\|}_{\underline{3},0}^2
+{\|\sqrt{\mu}\partial_3^3\eta\|}_{\underline{2},0}^2)\nonumber
\\
& \leqslant  \frac{\Lambda}{2}
  ({\|\sqrt{\mu}\partial_3^2\eta\|}_{\underline{3},0}^2
+{\|\sqrt{\mu}\partial_3^3\eta\|}_{\underline{2},0}^2 )+c ( \|(\eta,u)\|_{\underline{4},1}^2 + \|u_{tt}\|_1^2+\sqrt{ \mathfrak{E}} \mathfrak{D}).\nonumber
\end{align}
Using  \eqref{2017020614181721xx}--\eqref{201901031537}  and Young inequality, we derive from \eqref{201901031937} with  $2\leqslant j\leqslant 3$ that
\begin{align} &
\frac{\mm{d}}{\mm{d}t} ({\|\sqrt{\mu}\partial_3^4\eta\|}_{\underline{1},0}^2+
{\|\sqrt{\mu}\partial_3^5\eta\|}_{0}^2)\nonumber
\\
&\leqslant  \frac{\Lambda}{2}({\|\sqrt{\mu}\partial_3^4\eta\|}_{\underline{1},0}^2+
{\|\sqrt{\mu}\partial_3^5\eta\|}_{0}^2
)+c
\left( \|\eta\|_{ \underline{2},3}^2
 +
\|u\|_{\underline{2},1}^2 +\| u_{tt} \|_1^2 +\sqrt{ \mathfrak{E}} \mathfrak{D}\right).\nonumber
 \end{align}

We can derive from \eqref{201702061553} and the above two estimates that, for any sufficiently small $\delta_4$ and any sufficiently large $c_3$,
\begin{align}
\frac{\mm{d}}{\mm{d}t}\mathfrak{E}_3 + cc_3^2\mathfrak{O}_1 \leqslant &
 \frac{3\Lambda c_3}{4}\left(
  \| \sqrt{\mu}\partial_3^{2} \eta \|_{\underline{3},0}^2
 + \|\sqrt{\mu} \partial_3^3 \eta \|_{\underline{2},0}^2\right)\nonumber \\
  & + \frac{\Lambda}{2}({\|\sqrt{\mu}\partial_3^4\eta\|}_{\underline{1},0}^2+
{\|\sqrt{\mu}\partial_3^5\eta\|}_{0}^2 \nonumber
)\\
&  + \left( c_3^3(\|\eta\|_{\underline{4},0}^2+ \|u_3\|_0^2+ \sqrt{{\mathfrak{E}}} {\mathfrak{D}})+c_3\|\eta\|_{\underline{4},1}^2 \right)/c , \label{2018081315583}
\end{align}
where we have defined that
\begin{align}
\nonumber
& \mathfrak{E}_3:=  c_3^2\mathfrak{E}_1 + c_3 ( {\|\sqrt{\mu}\partial_3^2\eta\|}_{\underline{3},0}^2
+{\|\sqrt{\mu}\partial_3^3\eta\|}_{\underline{2},0}^2)+ {\|\sqrt{\mu}\partial_3^4\eta\|}_{\underline{1},0}^2+
{\|\sqrt{\mu}\partial_3^5\eta\|}_{0}^2.
 \end{align}
Making use of \eqref{2017020614181721}--\eqref{2017020614181721xxx}, \eqref{201702071610nb}, and Friedrichs's and Young's inequalities, we can derive that, there exists a constant $c$, such that, for any sufficiently small $\delta_4$ and any sufficiently large $c_3$,
\begin{align}
& \mathfrak{E}_3,\ \mathfrak{E}\mbox{ and } \|\eta\|_5^2+\|u\|_4^2 \mbox{ are equivalent}\mbox{ for any }t \geqslant 0,\\
\label{201808safdas02asdfsadf1219}
&c \left(c_3^2  \|\eta\|_{\underline{4},1}^2+\|\eta\|_5^2 \right) \leqslant  \mathfrak{E}_3,\\
\label{201808061629}&
\mathfrak{O}_1+  \|\eta\|_5\mbox{ is equivalent to }\mathfrak{D}.
\end{align}
Thus we easily further derive \eqref{2016121521430850}--\eqref{2018008121027}   from \eqref{2018081315583}--\eqref{201808061629}.\hfill $\Box$\end{pf}

Consequently,  similarly to Proposition \ref{pro:0401xx}, we have the following conclusion.
\begin{pro} \label{pro:0401nxd}
\begin{enumerate}
  \item[(1)] Let  $(\eta^0,u^0)\in H^{5,1}_{0,*}\times H_0^4$ and  $\zeta^0:=\eta^0+y$.
           There exists a sufficiently small ${\delta}_5\in (0,1)$, such that, if $(\eta^0,u^0)$ satisfying
\begin{align}
\label{201801032021}
& \sqrt{\|\eta^0\|_5^2+\|{u}^0\|_4^2}<  {\delta}_5
\end{align}
and necessary compatibility conditions
\begin{alignat}{2}
& \mm{div}_{\mathcal{A}^0}u^0=0& &\ \mbox{ in }\Omega_{\mm{p}}, \label{201808040977}\\
&\partial_tu|_{t=0}=0 & &\ \mbox{ in }\Omega_{\mm{p}} ,\label{201808040977xx}
\end{alignat}
then there is a  local existence time $T^{\max}>0$ (depending on ${\delta}_5$, the domain, the density profile and the known parameters), and a unique local-in-time classical  solution
$(\eta, u,\sigma)\in C^0([0,T^{\max}) ,  H^{5,1}_{0,*}\times H_0^4\times \underline{H}^3) $ to the transformed MRT problem, where the solution enjoys the regularity \eqref{201811142106}--\eqref{201811142107} with $[0, T^{\max})$, reps. $I_{T^{\max}}$, in place of $\overline{I_T}$, resp. $I_T$.
   \item[(2)] If the solution $(\eta,u)$  further satisfies
$$\sup_{t\in [0,T)}\sqrt{\|\eta(t)\|_5^2 +\|u(t)\|_4^2}\leqslant {\delta}_4\mbox{ for some }T<T^{\max},$$
 then  $(\eta, u)$ enjoys the Gronwall-type energy inequality \eqref{2016121521430850} and
 the equivalent estimate \eqref{2018008121027}.
\end{enumerate}
\end{pro}

\begin{rem}
For any given initial data $(\eta_0, {u}_0)\in H^{5,1}_{0,*}\times H^4_0$ satisfying  \eqref{201801032021}--\eqref{201808040977} with sufficiently small $\delta_5$, there exists  a unique local-in-time strong solution $(\eta, {u},\sigma)\in C^0([0,T),H^{3,1}_{0,*}\times H^2_0\times \underline{H}^1)$. Moreover, the initial date of  $\sigma$ is a weak solution to
\begin{equation}
\left\{
\begin{array}{ll}
\mm{div}_{\mathcal{A}^0}(( \nabla_{\mathcal{A}^0}\sigma^0 ) /\bar{\rho})=\mm{div}_{\mathcal{A}^0} \left(( \mu \Delta_{\mathcal{A}^0}u^0+\lambda\partial_{\bar{M}}^2\eta^0+gG^0e_3)/\bar{\rho}  - u^0 \cdot \nabla_{\mathcal{A}^0} u^0  \right)  &\mbox{ in }\Omega_{\mm{p}},\\
 \nabla_{\mathcal{A}^0}\sigma^0  \cdot   e_3 =\left(\mu\Delta_{\mathcal{A}^0}u^0+ \lambda \partial_{\bar{M}}^2\eta^0+G^0 \right) \cdot   e_3 &\mbox{ on }\partial\Omega_{\mm{p}},
\end{array}
\right. \label{201808040916}
\end{equation}
where $G^0:=\bar{\rho}(y_3+\eta_3^0)-\bar{\rho}(y_3)$.
If the condition \eqref{201808040977xx} is further satisfied, i.e., $(\eta^0,u^0,\sigma^0)$ satisfies
\begin{equation}
\label{201812271841x}
\nabla_{\mathcal{A}^0} \sigma^0-\mu\Delta_{\mathcal{A}^0} u^0- G^0-\lambda\partial_{\bar{M}}^2\eta^0=0 \mbox{ on }\partial\Omega_{\mm{p}},
\end{equation}
then we can improve the regularity of $(\eta,u,\sigma)$ so that it is a classical solution for sufficiently small ${\delta}_5$.

In addition, since there exists a constant ${\delta}_6$ such that, for any $\|w\|_3\leqslant {\delta}_6$
$$\|\nabla f\|_0 \leqslant c\|\nabla_{\mathcal{A}} f\|_0 \mbox{ for any }f\in H^1,$$
where $\mathcal{A}^{\mm{T}}:=(\nabla w)^{-1}$. Thus we easily see that, for any
$\|\eta^0\|_3\leqslant {\delta}_6$, the solution solution $\sigma^0\in \underline{H}^1$ to \eqref{201808040916} is unique  for given $(\eta^0,u^0)$.
\end{rem}

\begin{rem}
\label{201811031411}
Similarly to Remark \ref{201811031411xx},  for any classical solution $(\eta,u,\sigma)$ constructed by Proposition \ref{pro:0401nxd}, we take  $(\eta,u,\sigma)|_{t=t_0}$ as a new initial  datum,  where $t_0\in I_{T^{\max}}$. Then the new initial data can define a unique local-in-time classical solution $(\tilde{\eta},\tilde{u},\tilde{\sigma})$ constructed by Proposition
\ref{pro:0401nxd}; moreover the initial date of $\tilde{\sigma}$ is equal to $\sigma|_{t=t_0}$, if $\|\eta\|_3 \leqslant \delta_6 $ in $I_{T^{\max}}$.
\end{rem}


\subsection{Construction of initial data for  nonlinear solutions}

Let us recall the linear instability result of the transformed MRT problem \eqref{n0101nnxxs}.
\begin{pro}\label{thm:0201201622}
Under the assumptions of Theorem \ref{201806dsaff012301}, then zero solution
 is unstable to the linearized MRT problem
 \begin{equation*}   \left\{\begin{array}{ll}
\eta_t=u &\mbox{ in } \Omega_{\mm{p}},\\[1mm]
\bar{\rho} u_t+ \nabla \sigma-\Delta u=\lambda\partial_{\bar{M}}^2\eta +g\bar{\rho}'\eta_3 e_3&\mbox{ in }  \Omega_{\mm{p}},\\[1mm]
\mathrm{div} {u} =0& \mbox{ in }  \Omega_{\mm{p}},\\
 (\eta,u)=0 &\mbox{ on }\partial\Omega_{\mm{p}} ,
\end{array}\right.\end{equation*}
 That is, there is an unstable solution
$(\eta, u,  {\sigma}):=e^{\Lambda t}(-\bar{\rho}'\tilde{u}_3/\Lambda,\tilde{ {u}}, \tilde{\sigma})$
 to  the linearized MRT problem, where $(\tilde{u},\tilde{\sigma} )\in H^5_0\times \underline{H}^4$
 solves  the boundary value problem:
 \begin{equation*}
\left\{    \begin{array}{l}
   \Lambda^2\bar{\rho}\tilde{u} =\Lambda\mu\Delta\tilde{u}
-\Lambda \nabla \tilde{\sigma} +    \lambda
 \partial_{\bar{M}}^2 \tilde{u} +g\bar{\rho}'\tilde{u}_3e_3 , \;\;  \mathrm{div}\tilde{u}=0, \\
   \tilde{u}|_{\partial\Omega}= {0}  \end{array}  \right.
                       \end{equation*}
  with a (largest)
growth rate  $\Lambda>0$ satisfying
\begin{equation}\label{Lambdard}
\begin{aligned}
\Lambda^2=\sup_{{ {w}}\in H_\sigma^1 } \frac{\Lambda\mu\|\nabla {w}\|^2_0-E( {w})}{\int \bar{\rho}| {w}|^2\mm{d}y}. \end{aligned}\end{equation}
Moreover,
\begin{align}
\label{201602081445MH}
&\|\tilde{u}_{\mm{h}}\|_{L^1}\|\tilde{u}_3\|_{L^1} >0,\\
&\|\partial_{\bar{M}}u^0_{\mm{h}}\|_{L^1}\|\partial_{\bar{M}}\tilde{u}^0_3\|_{L^1}>0\mbox{ if }\bar{M}_3\neq 0.
\label{201602081445MHxx}\end{align}
\end{pro}
\begin{pf}  The linear instability result for the case $\bar{M}_{\mm{h}}=0$ is obvious by observing the proof of \cite[Proposition 5.1]{JFJSJMFMOSERT}.

For the case $\bar{M}_{\mm{h}}\neq 0$, we shall further use the following three assertions:
\begin{enumerate}
  \item[(1)] It holds that
$$  \sup_{0 \neq w\in H_\sigma^{1}  } \frac{ \int \bar{\rho}' w_3^2\mm{d}y}{ \|\partial_3 w\|^2_{0}}=\sup_{0 \neq w\in H_\sigma^{1}   } \frac{ \int \bar{\rho}' w_3^2\mm{d}y}{ \|\partial_{\bar{M}} w\|^2_{0}}>0,$$
see \cite[Lemma 4.6]{JFJSOMITN} for the proof. By the above identity and the instability condition \eqref{201812291920}, we see that $E( {w})<0$
for some $w\in H_\sigma^{1}$.
\item[(2)] If $\bar{M}_3\neq 0$, then $\partial_{\bar{M}}w=0$ for $w\in H^1_0$, if only if $w=0$ (see \cite[Remark  4.3]{JFJSOMITN}). This assertion will be used to derive \eqref{201602081445MHxx}.
  \item[(3)]  \emph{If $\tilde{u}\in H_\sigma^1$ is a weak solution to the boundary-value problem
 \begin{equation}
 \label{201901032056}
\left\{  \begin{array}{l}
 \nabla\tilde{\sigma} -\alpha\mu \Delta\tilde{u}
 = \lambda
 \partial_{\bar{M}}^2 \tilde{u} +g\bar{\rho}'\tilde{u}_3e_3 -\beta\bar{\rho}\tilde{u} , \quad \mathrm{div}\tilde{u}=0, \\
   \tilde{u}|_{\partial\Omega}= {0}   \end{array}  \right.
                       \end{equation}
for any given constants $\alpha>0$ and $\beta$.
 Then $\tilde{u}\in H_\sigma^5$ is a classical solution to the above boundary-value problem \eqref{201901032056} with  $\tilde{\sigma}\in \underline{H}^4$.}

We briefly mention how to deduce the above assertion for reader's convenience. Noting that $\tilde{u}$ satisfies
$$\alpha\mu\|\nabla \tilde{u}\|^2_0+\lambda \|\partial_{\bar{M}}^2\tilde{u}\|_0^2=\int (g\bar{\rho}'\tilde{u}_3e_3 -\beta\bar{\rho}\tilde{u})\cdot \tilde{u}\mm{d}y, $$
thus, by the above identity structure and a standard difference quotient method, we easily prove $\partial_{\mm{h}}^i \tilde{u} \in H^1_0$ for any $i\geqslant 1$. Using the classical regularity theory of Stokes problem, we easily obtain the desired assertion.
\end{enumerate}

Consequently, using the above three assertions, we easily obtain Proposition \ref{thm:0201201622} by directly following the argument of \cite[Proposition 5.1]{JFJSJMFMOSERT}.
\end{pf}

For any given $\delta>0$,  let
\begin{equation}\label{0501}
\left(\eta^\mm{a}, u^\mm{a} ,\sigma^{\mm{a}}\right)=\delta e^{{\Lambda t}} (\tilde{\eta}^0,\tilde{u}^0,\tilde{\sigma}^0),
\end{equation}
where $(\tilde{\eta}^0,\tilde{u}^0, \tilde{\sigma}^0):=(\tilde{u}/\Lambda,\tilde{u},\tilde{\sigma})$, and $(\tilde{u},\tilde{\sigma})\in  H^5_0 \times  \underline{H}^4$ comes from  Proposition \ref{thm:0201201622}.
Then $\left(\eta^\mm{a}, u^\mm{a}\right)$  is a solution to the linearized MRT problem, and enjoys the estimate, for any $i \geqslant 0$,
 \begin{equation}
\label{appesimtsofu1857}
\|\partial_t^i(\eta^{\mm{a}},u^{\mm{a}}, \sigma^{\mm{a}})\|_{\mm{S},2}\leqslant c(i) \delta e^{\Lambda t}.
 \end{equation}Moreover, by \eqref{201602081445MH} and \eqref{201602081445MHxx}, for $\tilde{\chi}^0:=\tilde{\eta}^0$ or $\tilde{u}^0$,
 \begin{align}
&  \|\tilde{\chi}_{\mm{h}}^0\|_{L^1} \|\tilde{\chi}_3^0\|_{L^1}
 > 0, \label{n05022052} \\
&\|\partial_{\bar{M}}\tilde{\chi}^0_{\mm{h}}\|_{L^1} \|\partial_{\bar{M}}\tilde{\chi}^0_3\|_{L^1}>0 \mbox{ if }
\bar{M}_3\neq0. \label{n05022052xx}
\end{align}
Next we modify  the initial data of the linear solutions.
\begin{pro}
\label{lem:modfied}
Let $(\tilde{\eta}^0,\tilde{u}^0,\tilde{\sigma}^0)$
be the same as in \eqref{0501}, then there exists a constant ${\delta}_7$, such that for any $\delta\in(0,{\delta}_7)$, there is a couple $(\eta^{\mm{r}},u^{\mm{r}},\sigma^{\mm{r}})\in H^5_0\cap H^4_0\cap \underline{H}^3$ enjoying the following properties:
\begin{enumerate}[\quad (1)]
                \item The modified initial data
  \begin{equation}\label{mmmode04091215}(  {\eta}_0^\delta,{u}_0^\delta,\sigma_0^\delta): =\delta
   (\tilde{\eta}^0,\tilde{u}^0, \tilde{\sigma}^0) + \delta^2 (\eta^\mm{r},   u^\mm{r},\sigma^{\mm{r}})
\end{equation}
belongs to $H^{5,1}_{0,*}\times H^4_0\times \underline{H}^3$, and satisfies
\begin{align}
\det \nabla ( {\eta}_0^\delta+y)=1,\nonumber
\end{align}
the compatibility conditions  \eqref{201808040977}, \eqref{201808040916}$_1$ and \eqref{201812271841x} with $(\eta_0^\delta,u_0^\delta,\sigma_0^\delta)$ in place of $(\eta^0,u^0$, $\sigma^0)$.
\item Uniform estimate:
\begin{equation}
\label{201702091755}
 \sqrt{\|\eta^{\mm{r}}\|_5^2+ \|(u^{\mm{r}},\sigma^{\mm{r}})\|_{\mm{S},2}^2 }\leqslant {C}_7,
 \end{equation}
where the constant ${C}_7\geqslant 1$ depends on the domain, the density profile and the known parameters, but is independent of $\delta$.
 \end{enumerate}
\end{pro}
\begin{pf}
Recalling the construction of $(\tilde{\eta}^0,\tilde{u}^0,\tilde{\sigma}^0)$, we can see that
$(\tilde{\eta}^0,\tilde{u}^0,\tilde{\sigma}^0)$ satisfies
\begin{equation}
\label{201705141510}
\left\{
\begin{array}{ll}
 \mm{div}\tilde{\eta}^0=\mm{div}\tilde{u}^0=0   &\mbox{ in }\Omega_{\mm{p}},\\
 \Lambda\bar{\rho} \tilde{u}^0 +\nabla\tilde{\sigma}^0-\mu\Delta \tilde{u}^0=\lambda\partial_{\bar{M}}^2\tilde{\eta}^0 +g\bar{\rho}'\eta_3 e_3  & \mbox{ in }\Omega_{\mm{p}},\\
 (\tilde{\eta}^0,\tilde{u}^0) = {0} & \mbox{ on }\partial\Omega_{\mm{p}}.
 \end{array}\right.
\end{equation}
If $(\eta^\mm{r},u^\mm{r},\sigma^\mm{r})\in H^5_0\times H^4_0\times \underline{H}^3$ satisfies
\begin{equation}
\label{201702061320}
\left\{\begin{array}{ll}
\mm{div}\eta^{\mm{r}}=O(\eta^{\mm{r}}),\ \mm{div}u^\mm{r} = \mm{div}(\tilde{\mathcal{A}}^{\mm{T}} u_0^\delta)/\delta^2 &\mbox{ in }\Omega_{\mm{p}},\\
 \nabla \sigma^{\mm{r}}-\mu\Delta u^{\mm{r}}= \lambda\partial_{\bar{M}}^2\eta^{\mm{r}}+g\bar{\rho}'\eta^{\mm{r}}_3e_3 + \bar{\rho} \Upsilon^M + \left(\mu(\nabla_{\tilde{\mathcal{A}}_0^\delta}\mm{div}_{\mathcal{A}_0^\delta}u^\delta_0+
 \nabla\mm{div}_{\tilde{\mathcal{A}}_0^\delta}u^\delta_0)\right. & \\
\left.-  \bar{\rho}u^\delta_0 \cdot \nabla_{\mathcal{A}^0_\delta} u^\delta_0
  -\nabla_{\mathcal{A}_0^\delta} \sigma^\delta_0 +g\int_0^{\eta_0^\delta(y,t)\cdot e_3}
\int_0^\tau\bar{\rho}''(y_3+s)\mm{d}s\mm{d}\tau e_3
  \right)/\delta^2  &\mbox{ in }\Omega_{\mm{p}},  \\
     \mm{div}_{\mathcal{A}_0^\delta}\Upsilon^M=- \delta \Lambda \mm{div}_{\tilde{\mathcal{A}}_0^\delta}\tilde{u}^0  &   \mbox{ in }\Omega_{\mm{p}},\\
(\eta^\mm{r}, u^\mm{r},\Upsilon^M )=0&\mbox{ on }\partial\Omega_{\mm{p}},
\end{array}\right.\end{equation}
 where $({\eta} ^\delta_0,{u} ^\delta_0,\sigma ^\delta_0)$ is given in the mode \eqref{mmmode04091215},
  $\zeta^\delta_0:=\eta^\delta_0 +y$, $\mathcal{A} ^\delta_0:= (\nabla\zeta^{\delta}_0)^{-\mm{T}}$, $\tilde{\mathcal{A}}^{\delta}_0 := \mathcal{A} ^\delta_0-I$ and $O(\eta^{\mm{r}}):=\delta^{-2}\mm{div}
\Psi(\delta\tilde{\eta}+\delta^2\eta^{\mm{r}})$,
see \eqref{201812091021} for the definition of $\Psi$.
Then, by \eqref{201812091021} and \eqref{201705141510}, it is easy to check that $({\eta} ^\delta,{u} ^\delta,\sigma ^\delta)$  belongs to $H^{5,1}_{0,*}\times H^4_0 \times \underline{H}^3$, and satisfies
\eqref{201808040977}, \eqref{201808040916}$_1$ and \eqref{201812271841x}  with $(\eta ^\delta,u ^\delta,\sigma ^\delta)$ in place of $(\eta^0 ,u^0 ,\sigma^0 )$.
Next we briefly introduce how to  construct such $(\eta^\mm{r},u^\mm{r},\sigma^\mm{r} )$ enjoying \eqref{201702061320}.

For sufficiently small $\delta_7$, following the argument of the construction of $(u^{\mm{r}},q^{\mm{r}})$ in Proposition \ref{lem:modfiedxx}, we can also construct a $(\eta^{\mm{r}}, \varpi)\in H^5_0\times \underline{H}^4$ such that, for any $\delta \leqslant \delta_7$,
\begin{equation}
\label{20170514xx}
 \left\{\begin{array}{ll}
\nabla \varpi- \Delta \eta^{\mm{r}} =0,\quad  \mm{div}\eta^{\mm{r}} =O(\eta^{\mm{r}}) &\mbox{ in }\Omega_{\mm{p}},  \\
\eta^{\mm{r}} =0&\mbox{ on } \partial\Omega_{\mm{p}}
\end{array}\right.  \end{equation}
and $\|\eta^{\mm{r}}\|_5\leqslant c$ for some constant $c$, which is independent of $\delta$.
 Similarly to \eqref{2019901122315}, we also have $\eta_0^\delta\in H^{5,1}_{0,*}$ for sufficiently small  $\delta_7$.

Following the construct of $\Upsilon$ in the proof of Proposition \ref{lem:modfiedxx}, we can also get a function $\Upsilon^M$ such that, for any sufficiently $\delta_7$ and $\eta^{\mm{r}}$ constructed by \eqref{20170514xx},
\begin{equation}
\label{201702051905xfwea}
 \left\{\begin{array}{ll}
\nabla q-  \Delta \Upsilon^M  =0, \
 \mm{div}\Upsilon^M=\mm{div}_{\tilde{\mathcal{A}}_0^\delta}(\Upsilon^M- \delta \Lambda \tilde{u}^0) &\mbox{ in }\Omega_{\mm{p}}, \\
\Upsilon^M=0 &\mbox{ on }\partial\Omega_{\mm{p}}
\end{array}\right.\end{equation}
and
\begin{equation}
\|(\Upsilon^M,q)\|_{\mm{S},0}\lesssim  \delta^2.
\label{xx}
\end{equation}

Finally, for sufficiently small $\delta_7$,  we can further construct $(u^{\mm{r}},\sigma^{\mm{r}})\in H^4_0\times \underline{H}^3$ such that, for any $\delta \leqslant \delta_7$,
\begin{equation}
\label{201702061320xx}
\left\{\begin{array}{ll}
 \nabla \sigma^{\mm{r}}-\mu\Delta u^{\mm{r}}= \lambda\partial_{\bar{M}}^2\eta^{\mm{r}}+g\bar{\rho}'\eta^{\mm{r}}_3e_3+ \bar{\rho} \Upsilon^M + \left(\mu(\nabla_{\tilde{\mathcal{A}}_0^\delta}\mm{div}_{\mathcal{A}_0^\delta}u^\delta_0+
 \nabla\mm{div}_{\tilde{\mathcal{A}}_0^\delta}u^\delta_0)\right.& \\
\left.-  \bar{\rho}u^\delta_0 \cdot \nabla_{\mathcal{A}^0_\delta} u^\delta_0
  -\nabla_{\mathcal{A}_0^\delta} \sigma^\delta_0 +g\int_0^{\eta_0^\delta(y,t)\cdot e_3}
\int_0^\tau\bar{\rho}''(y_3+s)\mm{d}s\mm{d}\tau e_3
  \right)/\delta^2  &\mbox{ in }\Omega_{\mm{p}},\\
\mm{div}u^\mm{r} =\mm{div}(\tilde{\mathcal{A}} u_0^\delta)/\delta^2 &\mbox{ in }\Omega_{\mm{p}},\\
    u^\mm{r} =0&\mbox{ on }\partial\Omega_{\mm{p}}
\end{array}\right.\end{equation}
and $\|u^{\mm{r}}\|_4\leqslant c$ for some constant $c$, which is independent of $\delta$.
Consequently, we finish the construction of $(\eta^\mm{r},u^\mm{r},\sigma^\mm{r} )$.  \hfill $\Box$
\end{pf}

\subsection{Error estimates and existence of escape times}

Let
$$
\begin{aligned}
&{C}_8:= \sqrt{\|\tilde{\eta}^0\|_5^2+ \|\tilde{u}^0\|_4^2 } +{C}_7 \geqslant 1,\\
 & \delta<{\delta}_0^M:= \min\{{\delta}_4,{\delta}_5,\delta_6, {C}_8{\delta}_7\}/2{C}_8<1,
 \end{aligned}$$
and $(\eta_0^\delta,u_0^\delta,q_0^\delta)$ be constructed by Proposition \ref{lem:modfied}.

 Noting that
\begin{align}
  \sqrt{\|\eta_0^\delta\|_5^2
+\|u_0^\delta\|_4^2 }\leqslant {C}_8\delta<2{C}_8{\delta}_0^M\leqslant {\delta}_5 ,
\label{201809121553}
\end{align}
by Proposition \ref{pro:0401nxd}, there exists a (nonlinear) solution $(\eta,u ,\sigma)$ of the transformed MRT problem defined on some time interval $I_{T^{\max}}$ with initial value $(  {\eta}_0^\delta ,{u}_0^\delta )$.

Let $\epsilon_0\in (0,1)$ be a constant, which will be defined in \eqref{defined}.
 We define
 \begin{align}\label{times}
& T^\delta:={\Lambda}^{-1}\mm{ln}({\epsilon_0}/{\delta})>0,\quad\mbox{i.e.,}\;
 \delta e^{\Lambda T^\delta }=\epsilon_0,\\
&T^*:=\sup\left\{t\in I_{T^{\max}} \left|~\sqrt{{\|\eta (\tau)\|_5^2+
\| u (\tau)\|_4^2 }}\right.\right. \leqslant 2{C}_8{\delta}_0^M\mbox{ for any }\tau\in [0,t)\bigg\},\nonumber\\
&  T^{**}:=\sup\left\{t\in  I_{T^{\max}} \left|~\left\|(\eta,u)(\tau)\right\|_{0}\leqslant 2 {C}_8\delta e^{\Lambda \tau}\mbox{ for any }\tau\in [0,t)\right\}.\right.\nonumber
 \end{align}
Then $T^*>0$ and $T^{**}>0$.

Noting that $2C_8\delta_0^M\leqslant \delta_6$, thus, similarly to \eqref{0502n1xx}--\eqref{0502n111},  we also have
 \begin{align}
&\sqrt{\|\eta (T^*)\|_5^2+
\| u (T^*)\|_4^2 }=2{C}_8 {\delta}_0^M,\mbox{ if }T^*<\infty ,\nonumber \\
& \left\|(\eta,u) (T^{**}) \right\|_0
=2 {C}_8\delta e^{\Lambda T^{**}},\mbox{ if }T^{**}<T^{\max}.\nonumber
\end{align}

We denote ${T}^{\min}:= \min\{T^\delta ,T^*,T^{**} \}$. Noting that $(\eta,u)$ satisfies
$$
 \sup_{0\leqslant t< T^{\min}}\sqrt{\|\eta(t)\|_5^2 +\|u(t)\|_4^2 }\leqslant {\delta}_4,$$
thus,  for any $t\in I_{T^{\min}}$, $(\eta,u)$ enjoys  \eqref{2016121521430850} and \eqref{2018008121027} with $(\eta^\delta_0,u^\delta_0)$ in place of $(\eta^0,u^0)$ by the second conclusion in Proposition \ref{pro:0401nxd}.
Noting that $\|(\eta,u)(t)\|_0\leqslant 2{C}_8 \delta e^{\Lambda t}$ on $I_{T^{\min}}$, then we deduce from the estimate \eqref{2016121521430850} and \eqref{201809121553}  that, for all $t\in I_{T^{\min}}$,
\begin{align}
  \tilde{\mathfrak{E}}(t)  \leqslant c \delta^2e^{2\Lambda t}+\Lambda\int_0^t \tilde{\mathfrak{E}}(\tau)  \mm{d}\tau.
 \label{0503} \end{align}
Applying the Gronwall's inequality to the above estimate  yields
$\tilde{\mathfrak{E}}(t)\lesssim (\delta e^{\Lambda t})^2$, which, together with \eqref{2018008121027}, further implies
\begin{align}
& \mathfrak{E}(t) \leqslant ({C}_9\delta e^{\Lambda t})^2\leqslant {C}_9^2\epsilon_0^2\mbox{ in } I_{T^{\min}}.
\label{201702092114}
\end{align}

Next we estimate the error between the (nonlinear) solution $(\eta,u )$ and the linear solution $(\eta^\mm{a},u^{\mm{a}})$ provided by \eqref{0501}.
\begin{lem}\label{lem:0401xxxx}
  There exists a constant ${C}_{10}$ such that,  for any $t\in I_{T^{\min}}$,
\begin{align}
\label{ereroe}
&\|(\eta^{\mm{d}}, u^{\mm{d}})\|_{\mathfrak{X}}  \leqslant C_{10}\sqrt{\delta^3e^{3\Lambda t}},
\end{align}
where $(\eta^{\mathrm{d}}, u^{\mathrm{d}}):=(\eta, u)-(\eta^\mm{a},u^{\mm{a}})$, $ \mathfrak{X}=W^{1,1}$ or $H^1$, and ${C}_{10}$ is independent  of  $\delta$,  and $T^{\min}$.
\end{lem}
\begin{pf}
Subtracting the both transformed MRT  and the linearized MRT problems, we get
\begin{equation}\label{201702052209}\left\{\begin{array}{ll}
\eta_t^{\mathrm{d}}=u^{\mathrm{d}} &\mbox{ in } \Omega_{\mm{p}},\\[1mm]
\bar{\rho} u_{t}^{\mathrm{d}}+\nabla \sigma^{\mm{d}}- \mu\Delta u^{\mm{d}} - g\bar{\rho}'\eta_3^{\mm{d}}e_3
-\lambda \partial_{\bar{M}}^2 u^{\mm{d}}= \mathcal{N}^3& \mbox{ in }\Omega_{\mm{p}}, \\
\div_{\mathcal{A}}  u=0& \mbox{ in }\Omega_{\mm{p}},\\
 (\eta^{\mathrm{d}},u^{\mathrm{d}})=0 &\mbox{ on }\partial\Omega_{\mm{p}},\\
 (\eta^{\mathrm{d}},u^{\mathrm{d}})|_{t=0}=\delta^2 (\eta^{\mm{r}},u^{\mm{r}}) &\mbox{ on }\Omega_{\mm{p}}.\end{array}\right.\end{equation}
Applying $\partial_t $ to \eqref{201702052209}$_2$--\eqref{201702052209}$_4$ yields that
\begin{equation}\label{201702052221}\left\{\begin{array}{ll}
  \bar{\rho} u_{tt}^{\mathrm{d}}+\nabla_{\mathcal{A}}\sigma^{\mathrm{d}}_t
-\Delta_{\mathcal{A}} u^\mm{d}_t&\\
=\lambda \partial_{\bar{M}}^2 u^{\mm{d}}+
g\bar{\rho}'u_3^{\mm{d}}e_3+ \mu( \mathrm{div}\nabla_{\tilde{\mathcal{A}}}u^{\mm{a}}_t
+ \mm{div}_{\tilde{\mathcal{A}}} \nabla_{\mathcal{A}}u^{\mm{a}}_t)-\nabla_{\tilde{\mathcal{A}}}\sigma^{\mm{a}}_t+\mathcal{N}^4 & \mbox{ in }\Omega_{\mm{p}}, \\
\div_{\mathcal{A}}  u^{\mathrm{d}}_t =-\mm{div}_{\mathcal{A}_t}u- \mm{div}_{\tilde{\mathcal{A}}}u^{\mm{a}}_t  & \mbox{ in }\Omega_{\mm{p}},\\
 u^{\mathrm{d}} _t=0 &\mbox{ on }\partial\Omega_{\mm{p}}.
\end{array}\right.\end{equation}

Following the argument of \eqref{Lem:030dsfafds1m0dfsf832},
we can deduce from \eqref{201702052221} that
\begin{align}
& \frac{1}{2}\frac{\mm{d}}{\mm{d}t}\left(\|\sqrt{\bar{\rho}}  u_{t}^{\mm{d}}\|^2_0+ {E}( u^{\mm{d}}) \right)\mm{d}y + {\mu}\| \nabla_{\mathcal{A}}  u_{t}^{\mm{d}} \|_{0}^2\nonumber \\
&= \int \left(\mu \left( \mathrm{div} \nabla_{\tilde{\mathcal{A}}}u^{\mm{a}}_t +
 \mm{div}_{\tilde{\mathcal{A}}}\nabla_{\mathcal{A}}u_t^{\mm{a}}\right)
-\nabla_{\tilde{\mathcal{A}}}  \sigma^{\mm{a}}_t+\mathcal{N}^4\right)\cdot u_t^{\mm{d}}
\mm{d}y\nonumber  \\
& \quad
+\int (\nabla \sigma^{\mm{d}}_t  \cdot \mathcal{A}_t^{\mm{T}}u + \nabla \sigma^{\mm{d}}_t \cdot  (\tilde{\mathcal{A}}^{\mm{T}}u^{\mm{a}}_t ))\mm{d}y   =:R_1(t). \nonumber
\end{align}
Integrating the above identity in time from $0$ to $t$ yields that
\begin{align}
& \|\sqrt{\bar{\rho}} u_t^\mm{d}\|^2_{0}+E(u^{\mm{d}})+2 \mu\int_0^t\|\nabla_{\mathcal{A} } u^{\mathrm{d}}_t\|_0^2\mm{d}\tau  =2 \int_0^t {R}_1(\tau)\mm{d}\tau  + {E}(u^{\mm{d}}|_{t=0})+ \| \sqrt{\bar{\rho}}u_t^{\mathrm{d}}\big|_{t=0}\|^2_0 . \label{0314}
\end{align}

Using \eqref{appesimtsofu1857} and \eqref{201702092114},  we  easily estimate that
\begin{align}
\int_0^t {R}_1(\tau)\mm{d}\tau  \lesssim   \delta^3 e^{3\Lambda t}  . \label{201811081020}
\end{align}
Noting that $u^\mm{d}(0)=\delta^2 u^{\mm{r}}$, we have
\begin{equation}
E(u^{\mm{d}}|_{t=0})\lesssim \delta^4\|u^{\mm{r}}\|_2^2\lesssim  \delta^3 e^{3\Lambda t}.
\label{2018110702038}
\end{equation}

Next we turn to estimate for $\| \sqrt{\rho}u_t^{\mathrm{d}}\|^2_0\big|_{t=0} $.
Recalling that
$$\div  u_t^{\mm{d}}=  -\mm{div}\partial_t( \tilde{\mathcal{A}}^{\mm{T }}  u),$$
thus, multiplying  \eqref{201702052209}$_2$ by $u_t^{\mm{d}}$ in $L^2$, and using the integration by parts,  we have
\begin{align}
\int \bar{\rho} |u_{t}^{\mathrm{d}}|^2\mm{d}y=
  \int\bigg( \mu \Delta u^{\mm{d}}+\lambda\partial_{\bar{M}}^2\eta^{\mm{d}}+
g\bar{\rho}'\eta_3^{\mm{d}}e_3 +\mathcal{N}^3 \bigg)\cdot u_t^{\mm{d}}   \mm{d}y +
\int  \nabla  \sigma^{\mathrm{d}}\partial_t({\tilde{\mathcal{A}}}^{\mm{T}} u)  \mm{d}y  .\nonumber
  \end{align}
Thus we further  derive from the above estimate that $
  \|  u_t^{\mathrm{d}}\|_0^2 \lesssim\| (\eta^{\mm{d}},u^{\mm{d}})\|_2^2+ \delta^3 e^{3\Lambda t}$,
which, together with the initial data \eqref{201702052209}$_5$, implies that
\begin{equation}
\label{201807112050}
\|  u_t^{\mathrm{d}}|_{t=0}\|_0^2\lesssim  \delta^3 e^{3\Lambda t} .
\end{equation}

Consequently, putting \eqref{201811081020}--\eqref{201807112050} into \eqref{0314}  yields
\begin{align}
 \|\sqrt{\bar{\rho}} u_t^\mm{d}\|^2_{0}+ 2\mu\int_0^t\| \nabla_{\mathcal{A} } u^{\mathrm{d}}_{\tau}\|_0^2\mm{d}\tau\leqslant - {E}( u^{\mm{d}})+  c\delta^3 e^{3\Lambda t}.\label{0314xxdfafdss}
\end{align}

Since $u^{\mm{d}}$ does not satisfies the divergence-free condition (i.e.,
 $\mm{div}u^{\mm{d}}=0$), thus we can not use \eqref{Lambdard} to deal with $ -E(u^{\mm{d}})$.
To overcome this trouble, we consider the following  Stokes problem
\begin{equation*}
 \left\{\begin{array}{ll}
\nabla \varpi-\Delta \tilde{u}=0,\quad
 \mm{div}\tilde{u} =-  \div_{\tilde{\mathcal{A}}}
  u  &\mbox{ in }\Omega_{\mm{p}},  \\
\tilde{u}   =0&\mbox{ on } \partial\Omega_{\mm{p}}.
\end{array}\right.\end{equation*}
Then there exists a solution $(\tilde{u},\varpi)\in H_0^2\times  \underline{H}^1$ to the above Stokes problem for given $(\eta,u)$. Moreover,
\begin{equation}
\label{201705012219xx}
\|\tilde{u}\|_2\lesssim \|\mm{div}_{\tilde{\mathcal{A}}}u\|_1 \lesssim \delta^2 e^{2\Lambda t}.
\end{equation}
It is easy to check that $v^{\mm{d}}:=u^{\mm{d}}-
\tilde{u}\in H_\sigma^2$.

Now, we can apply \eqref{Lambdard} to $ -E(v^{\mathrm{d}})$, and get
  \begin{equation*}\begin{aligned}
-E(v^{\mathrm{d}})\leqslant  \Lambda^2{\|v^{\mathrm{d}}\|^2_0}+ \Lambda\mu\| \nabla  v^{\mathrm{d}}\|_0^2  ,
\end{aligned}\end{equation*}
which, together with \eqref{201705012219xx}, immediately implies
\begin{equation*}\begin{aligned}
-E(u^{\mathrm{d}})\leqslant  \Lambda^2{\|\sqrt{\bar{\rho}}u^{\mathrm{d}}\|^2_0} +  \Lambda{\mu}\|\nabla   u^{\mathrm{d}}\|_0^2 +c\delta^3 e^{3\Lambda t}.
\end{aligned}\end{equation*}
Inserting it into \eqref{0314xxdfafdss}, we arrive at
\begin{equation}
\label{201702232221}
\begin{aligned}  \| \sqrt{\bar{\rho}}u_t^\mm{d}\|^2_{0}+2\mu\int_0^t\|\nabla_{\mathcal{A}}
u^{\mathrm{d}}_{\tau}\|_0^2\mm{d}\tau \leqslant {\Lambda^2} \|  \sqrt{\bar{\rho}}u^\mm{d}\|_{0}^2 +{\Lambda} \mu\|\nabla u^{\mathrm{d}}\|_0^2 +  c \delta^3 e^{3\Lambda t}.
\end{aligned}\end{equation}
In addition,
$$
 \int_0^t\|\nabla u_{\tau}^{\mathrm{d}}\|_0^2\mm{d}\tau
\leqslant   \int_0^t(\| \nabla _{\mathcal{A}}u_{\tau}^{\mathrm{d}}\|_0^2+
\|\tilde{\mathcal{A}}\|_2\|u_{\tau}^{\mathrm{d}}\|_1^2)\mm{d}\tau
\leqslant \int_0^t\|\nabla_{\mathcal{A}}u_{\tau}^{\mathrm{d}}\|_0^2\mm{d}\tau+  c \delta^3 e^{3\Lambda t}. $$
Putting the above estimate into \eqref{201702232221} yields that
  \begin{equation} \label{new0311}
\begin{aligned}  \| \sqrt{\bar{\rho}} u_t^\mm{d}\|^2_{0}+2 \mu\int_0^t\| \nabla u^{\mathrm{d}}_{\tau}\|_0^2\mm{d}\tau \leqslant {\Lambda^2} \|\sqrt{\bar{\rho}} u^\mm{d}\|_{0}^2 +  \Lambda\mu  \|\nabla  u^{\mathrm{d}}\|_0^2+  c \delta^3 e^{3\Lambda t}.
\end{aligned}\end{equation}
Then we can further deduce from the above estimate that
\begin{eqnarray}\label{uestimate1n}
\|u^{\mathrm{d}}\|_{1 }^2+\| u_t^{\mathrm{d}}\|^2_0  \lesssim \delta^3e^{3\Lambda t}.
\end{eqnarray}

We turn to derive the error estimate for $\eta^{\mathrm{d}}$.
It follows from \eqref{201702052209}$_1$ that
\begin{equation*}\begin{aligned}
 \frac{\mm{d}}{\mm{d}t}\|\eta^{\mathrm{d}}\|_{1}^2
\lesssim   \| u^{\mathrm{d}}\|_1 \|\eta^{\mathrm{d}}\|_{1}.
\end{aligned}\end{equation*}
Therefore, using \eqref{uestimate1n} and  the initial condition ``$\eta^{\mm{d}}|_{t=0}=\delta^2 \eta^{\mm{r}}$",  it follows that
\begin{equation}\begin{aligned}\label{erroresimts}
 \|\eta^{\mathrm{d}}\|_{1 }\lesssim  &  \int_0^t \|u^{\mathrm{d}}\|_{1} \mm{d}\tau
+\delta^2\|\eta^{\mm{r}}\|_1\lesssim  \sqrt{ \delta^3e^{3\Lambda t}}.
\end{aligned}\end{equation}
Noting that $H^1\hookrightarrow W^{1,1}$, then we can derive
 \eqref{ereroe}  from \eqref{uestimate1n} and  \eqref{erroresimts}.
This completes the proof of Lemma  \ref{lem:0401xxxx}. \hfill$\Box$
\end{pf}

We define
 $$
\begin{aligned}
\varpi_0 :=& \left\{
         \begin{array}{ll}
\min_{\{\tilde{\chi}^0=\tilde{\eta}^0, \tilde{u}^0\}} \{ \|\tilde{\chi}^0_{\mm{h}}\|_{L^1},\|\tilde{\chi}^0_3\|_{L^1} \}, & \hbox{ if }\bar{M}_3=0;  \\
    \min_{\{\tilde{\chi}^0=\tilde{\eta}^0, \tilde{u}^0\}} \{ \|\tilde{\chi}^0_{\mm{h}}\|_{L^1},\|\tilde{\chi}^0_3\|_{L^1}, \|\partial_{\bar{M}} \tilde{\chi}^0_{\mm{h}}\|_{L^1}, \|\partial_{\bar{M}} \tilde{\chi}^0_3\|_{L^1} \}     & \hbox{ if }\bar{M}_3\neq 0;
         \end{array}
       \right.
\end{aligned}$$
By \eqref{n05022052} and \eqref{n05022052xx}, $\varpi_0>0$.
Now we define that
 \begin{equation}\label{defined}
\epsilon_0:=\min\left\{
 \frac{C_8\delta_0^M}{  C_9},
\frac{C_8^2}{4 C_{10}^2},
\frac{\varpi_0^2}{4C_{10}^2}   \right\}>0,
 \end{equation}
 Consequently, following the argument of the content after Lemma \ref{lem:0401xx}, we easily get  the  relation ``$T^\delta =T^{\min}\ne T^*$ or $T^{**}$" and the desired conclusion in Theorem \ref{thmsdfa3}.

\subsection{Proof of Theorem \ref{201806dsaff012301}}

Now we briefly introduce how to get Theorem \ref{201806dsaff012301} from Theorem \ref{thmsdfa3}. Let $(\eta,u)$ be an unstable solution constructed by Theorem \ref{thmsdfa3}. Since $\eta \in H_{0,*}^{5,1}$ for each $t\in \overline{I_T}$, then
 \begin{align}&
 {\zeta}   :\overline{\Omega_0^h}\to  \overline{\Omega_0^h} \mbox{ is a }C^0\mbox{-homeomorphism mapping},\label{201808151856sdfaxsfddx}  \\
& {\zeta} : \Omega_0^h \to {\zeta} ( {\Omega_0^h})\mbox{ is a }C^3\mbox{-diffeomorphic mapping},\label{201808151856sdfaxx}
\end{align}
Thus we can define that
\begin{align}
& \label{usedopso}
 \varrho:=\bar{\rho}({\zeta}^{-1}_3(x,t))-\bar{\rho}(x_3),\  v:=u|_{y=\zeta^{-1}(x,t)},\  N:= \partial_{\bar{M}}\eta|_{y=({\zeta}^{-1}(x,t))},\  q:=\sigma|_{y=\zeta^{-1}(x,t)} ,
\end{align}
where  $ \zeta (y,t): =\eta(y,t)+y$.
It is easy to check that $\varrho$, $u$, $N$ and $q$ are horizontally periodic functions as well as  $\eta$, $u$ and $\sigma$. Moreover, we have the following conclusion:
\begin{lem}
\label{201901151906}
Let $f\in L^1$,  $w\in    H_{0,*}^{3,1}$, $\varpi:=w+y$ and $\tilde{f}:=f(\varpi^{-1})$, then
$\|f\|_{L^1}=\|\tilde{f}\|_{L^1} $.
\end{lem}
\begin{pf}
Let $\mathcal{C}:=2\pi L_1\times 2\pi L_2\times (0,h)$ and $\varpi( {\mathcal{C}}):=\{x\in \mathbb{R}^3~|~x=\varpi(y),\ y\in  {\mathcal{C}}\}$.
Since $\tilde{f}$ is a horizontally periodic function and  $w\in    H_{0,*}^{3,1}$, we can check that
$$
\begin{aligned}
 \|f\|_{L^1}=& \int_{\varpi( {\mathcal{C}})}\tilde{f}\mm{d}x=\int_0^h
\int_{\{x_3=s\}\cap \varpi( {\mathcal{C}})}\tilde{f}(x)\mm{d}x_{\mm{h}}\mm{d}s\\
=& \int_0^h
\int_{2\pi L_1\times 2\pi L_2  }\tilde{f}(x_{\mm{h}},x_3)\mm{d}x_{\mm{h}}\mm{d}x_3= \|\tilde{f}\|_{L^1}.
\end{aligned}$$
This completes the proof. \hfill $\Box$
\end{pf}

By the regularity of $(\eta,u)$ and the embedding $H^{k+2}\hookrightarrow C^k(\overline{\Omega_{\mm{p}}})$ for $k\geqslant 0$,
$\tilde{\zeta} \in  C^0(\overline{\Omega ^T})\cap C^2(\Omega^T)$, where
$\tilde{\zeta} :=\tilde{\zeta} (y,t):=(\zeta (y,t),t)$ and $\Omega^T:=\Omega_0^h\times I_T$.
Moreover, by \eqref{201808151856sdfaxsfddx} and \eqref{201808151856sdfaxx}, we also have
\begin{align}&
\tilde{\zeta} (y,t) :\overline{\Omega ^T}\to  \overline{\Omega ^T} \mbox{ is a }C^0\mbox{-homeomorphism mapping}, \nonumber \\
&\tilde{\zeta} (y,t): \Omega ^T \to \tilde{\zeta} ( {\Omega ^T})\mbox{ is a }C^2\mbox{-diffeomorphic mapping}.\nonumber
\end{align}
In addition, by chain rule, we can compute out that
 \begin{equation}
\label{2018060321637}
\partial_t \zeta^{-1} =-((\nabla_y \zeta)^{-1}u )|_{y=\zeta^{-1} }=-(\mathcal{A}^{T} u )|_{y=\zeta^{-1} }
\end{equation}
and
 \begin{equation}
\label{2018060321637xx}
\nabla_x \zeta^{-1} =( \nabla \zeta)^{-1}|_{y=\zeta^{-1} },
\end{equation}
where $\mathcal{A}:=(\nabla \zeta )^{-\mm{T}}$.

By the regularity of $(\eta,u)$ and chain rule,
\begin{align}
\partial_{x_i}v= (\mathcal{A}_{i j}\partial_{ j} u)|_{y=\zeta^{-1}}.\label{201901092010}
\end{align}
Thus we can derive from \eqref{2018060321637} and \eqref{201901092010} that
\begin{align}
\label{2019010920044}
 v_t=(u_t-u\cdot\nabla_{\mathcal{A}}u)|_{y=\zeta^{-1}} .
 \end{align}
Making use of \eqref{201901092134},  \eqref{201901092145xx}, \eqref{2018060321637}, \eqref{201901092010}, \eqref{2019010920044}, Lemma \ref{201901151906}, the regularity of $(\eta,u)$ and chain rule, we easily derive that
\begin{align}
&(v,v_t,v_{tt})\in C^0( \overline{I_T}, H^4_\sigma\times H^2_\sigma\times L^2 )\cap L^2( I_T, H^5_\sigma\times H^3_\sigma\times H^1_\sigma),\nonumber \\
&
\mm{div}v=(\mm{div}_{\mathcal{A}}u)|_{y=\zeta^{-1}}=0, \|   v^0 \|_{4} \leqslant c\delta \mbox{ and } \eqref{201809041319},\nonumber
\end{align}
where $c$ is independent of $\delta$.

Similarly, by  the definition of $(\varrho,N,q)$, we also have
\begin{align}
& (q,q_t)\in C^0( \overline{I_T}, \underline{H}^3\times \underline{H}^1)\cap L^2( I_T, \underline H^4 \times \underline{H}^2),\ (\varrho,N)\in C^0( \overline{I_T}, H^5\times H^4) ,\nonumber \\
& \| {\varrho}^0\|_5+\| N^0\|_{4}+\|q ^0\|_{3}\leqslant c\delta\mbox{ and }\eqref{201809041319xx}.\nonumber\end{align}
Moreover, \begin{align}
&   \varrho_t=-{  v}\cdot\nabla (\varrho+\bar{\rho}), \nonumber \\
 & N_t:= (N+\bar{M})\cdot \nabla v- v\cdot\nabla (N+\bar{M}) , \ \mm{div}N=0,\nonumber  \\
 &(N+\bar{M})\cdot \nabla (N+\bar{M})=(\partial_{\bar{M}}^2\eta)|_{{y=\zeta^{-1}} }. \label{2019011302013}
\end{align}

Since $(\eta,\sigma)$ satisfies \eqref{n0101nnxxs}$_2$, then $(\varrho,v,N,q)$ solves \eqref{0103xx} by using \eqref{2019010920044}, \eqref{2019011302013} and chain rule.
 Hence $(\varrho,v,N,q)$ is a classical solution of the MRT problem \eqref{0103xx}--\eqref{0105xxx}. Moreover, the uniqueness of classical solution to the MRT problem can be verified by a standard energy method.
Finally, the compatibility conditions \eqref{20180923221611xx}--\eqref{20180923221611} directly follows
\eqref{20180923221611xx123}--\eqref{20180923221611xx1234}. This completes the proof of Theorem \ref{201806dsaff012301}.

\vspace{4mm} \noindent\textbf{Acknowledgements.}
 The research of Fei Jiang was supported by NSFC (Grant No. 11671086).

\renewcommand\refname{References}
\renewenvironment{thebibliography}[1]{%
\section*{\refname}
\list{{\arabic{enumi}}}{\def\makelabel##1{\hss{##1}}\topsep=0mm
\parsep=0mm
\partopsep=0mm\itemsep=0mm
\labelsep=1ex\itemindent=0mm
\settowidth\labelwidth{\small[#1]}%
\leftmargin\labelwidth \advance\leftmargin\labelsep
\advance\leftmargin -\itemindent
\usecounter{enumi}}\small
\def\newblock{\ }
\sloppy\clubpenalty4000\widowpenalty4000
\sfcode`\.=1000\relax}{\endlist}
\bibliographystyle{model1b-num-names}

\end{CJK*}
\end{document}